\documentclass[a4paper]{article}
\usepackage{graphicx}
\usepackage{amsmath}
\usepackage{amssymb}

\begin{document}
\newcommand{\bin}[2]{\left(\begin{array}{c}\!#1\!\\\!#2\!\end{array}\right)}
\newcommand{\threej}[6]{\left(\begin{array}{ccc} #1 & #2 & #3 \\ #4 & #5 & #6 \end{array}\right)}

\huge

\vspace{3cm}

\begin{center}
Accounting for highly excited states in detailed opacity calculations
\end{center}

\vspace{0.5cm}

\large

\begin{center}
Jean-Christophe Pain\footnote{jean-christophe.pain@cea.fr (corresponding author)} and Franck Gilleron
\end{center}

\vspace{0.2cm}

\normalsize

\begin{center}
CEA, DAM, DIF, F-91297 Arpajon, France
\end{center}

\vspace{0.5cm}


\begin{center}
{\bf Abstract}
\end{center}

In multiply-charged ion plasmas, a significant number of electrons may occupy high-energy orbitals. These ``Rydberg'' electrons, when they act as spectators, are responsible for a number of satellites of X-ray absorption or emission lines, yielding a broadening of the red wing of the resonance lines. The contribution of such satellite lines may be important, because of the high degeneracy of the relevant excited configurations which give these large Boltzmann weights. However, it is in general difficult to take these configurations into account since they are likely to give rise to a large number of lines. We propose to model the perturbation induced by the spectators in a way similar to the Partially-Resolved-Transition-Array approach recently published by C. Iglesias. It consists in a partial detailed-line-accounting calculation in which the effect of the Rydberg spectators is included through a shift and width, expressed in terms of the canonical partition functions, which are key-ingredients of the Super-Transition-Arrays model. The resulting method can \emph{a priori} be used in any detailed-configuration/line-accounting opacity code. 

\begin{center}
{\bf Keywords}
\end{center}

\begin{center}
hot plasma - atomic physics - highly excited states - satellite lines
\end{center}

\begin{center}
{\bf PACS} 
\end{center}

\begin{center}
32.70.Cs, 32.70.-n, 32.80.Ee, 32.80.Zb
\end{center}

\section{Introduction}\label{sec1}

\begin{figure}
\vspace{10mm}
\begin{center}
\includegraphics[width=8cm]{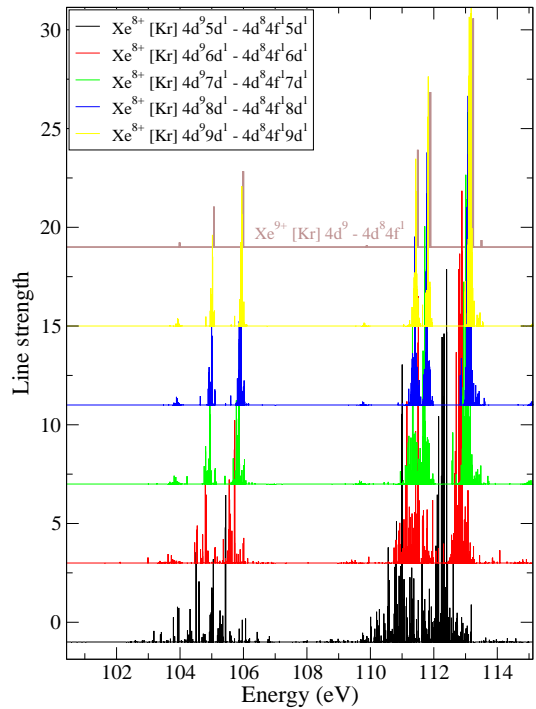}
\vspace{5mm}
\end{center}
\caption{(Color online) Effect of passive subshell nd$^1$, $n$=5 to 9, on transition array [Kr] 4d$^9$ $\rightarrow$ 4d$^8$4f$^1$. The calculation was performed with Cowan's code (RCN/RCN2/RCG).\label{Xe_bis}}
\end{figure}

Highly charged ions may have electrons in their high-lying loosely-bound orbitals. These ``Rydberg'' electrons, when they act as spectators, yield satellite lines that may not be clearly separated from the resonance line: they contribute as a broadening of its red wing. Some satellite lines are sufficiently strong that they can be used for diagnostics purposes. In ICF (Inertial Confinement Fusion) plasmas, electron temperature $T$ and density $N_e$ of the fuel are inferred from the X-ray emission of H-like and / or He-like lines of impurity ions, \emph{e.g.}, Ar ions \cite{KAWAMURA99}), which are enhanced and broadened by unresolved satellite lines. Concerning the Hohlraum emission, satellites of 3d-4f transitions of Ni-like gold ions form a quasi-continuum in the 2-4 keV energy range \cite{BAUCHE86}. In another context, a large number of satellite lines due to 4d-4f transitions contribute significantly to the EUV emission of Xe over the range [10-17] nm \cite{SASAKI04} (see Fig. \ref{Xe_bis}), which is important for lithographic applications.

The contribution of dielectronic satellite transition arrays, with spectator electrons in shell $n$=5, was studied by Bauche-Arnoult \emph{et al.} for the case of a tantalum plasma \cite{BAUCHE89}. The main mechanism for the formation of doubly-excited states is the dielectronic recombination. However, for high electron densities, those excited states can also be populated by successive collisional and radiative excitations from the inner shells. In general, the study of the corresponding satellite lines requires a non-LTE (local thermodynamic equilibrium) modeling, in which all microscopic processes are accounted for in a master equation in order to determine the population of states. Most of the doubly-excited states involved lie above the ionization potential of the next ion, so that they may broaden some transition arrays by autoionization. Biedermann \emph{et al.} \cite{BIEDERMANN97} measured the X-ray emission from high-$n$ spectator electrons following radiative stabilization in dielectronically excited Ba$^{40+}$ to  Ba$^{42+}$ ions produced by an electron beam ion trap. They observed an enhanced emission due to transitions from $n>$3 to $n$=3. 

In LTE, the contribution of those satellites may be important because of the large Boltzmann statistical weight of the relevant excited configurations (due to high electron temperature and/or high degeneracy). In fact, even if their individual probabilities are rather small, a large number of excited states may fill gaps in the spectrum \cite{BACK97}. In general, however, it is difficult to obtain a detailed description of those satellites because the number of electron configurations may become huge and they can give rise to a large number of lines (especially when open subshells with large angular momentum $\ell$ are involved). In that case, detailed-line-accounting (DLA) calculations become intractable.

A short review of semi-empirical and approximate models for treating high-$n$ spectators is presented in section \ref{semi}. These methods were proposed to address the aforementioned issue in a pragmatic manner. In section \ref{scorcg}, the detailed opacity code SCO-RCG  \cite{PORCHEROT11,PAIN13} is briefly described. The statistical approach that is currently used in the code to model those Rydberg spectators is explained. In section \ref{test}, we show the limit of this approach by studying the F-like 2p$\rightarrow$ 4d transition of an iron plasma at $T$=182 eV and $N_e=3.1~10^{22}$ cm$^{-3}$. These conditions correspond to a recent experiment performed by J. Bailey on the Sandia Z machine \cite{BAILEY15}. In section \ref{new}, we propose a new approach which allows one to improve the modeling of Rydberg spectators in the opacity code.
 
\section{Semi-empirical approaches and approximate models}\label{semi}

In Detailed-Configuration-Accounting codes, the Rydberg electrons can be only taken into account partially or approximately, because they are potentially responsible for a dramatic increase of the number of electron configurations. Thus, it is in general difficult to take those satellites into account, as they give rise to a large number of lines. In previous applications these contributions are either partially discarded, or described semi-empirically. Over the years, semi-empirical approaches and approximate models were proposed to solve that problem. In the middle of the eighties, Busquet \emph{et al.} introduced a broad empirical red wing \cite{BUSQUET85,BUSQUET00} in order to reproduce the M-shell gold spectra in the presence of myriad unresolvable satellite lines. At the same period, Goldberg and Rozsnyai \cite{GOLDBERG86} proposed to add a Gaussian width to the lines, based on statistical fluctuations of the populations of spectator electrons in high-lying states. Such an approach does not reproduce the red shift but accounts approximately for the way Rydberg spectators fill gaps between clusters of lines \cite{ROSE92}. The correction to the variance reads (see Appendix):

\begin{equation}\label{diel}
v_{\mathrm{dielectronic}}=\sum_{p~\mathrm{Rydberg}}\left(V_{\beta p}-V_{\alpha_p}\right)^2\bar{N}_p\left(1-\frac{\bar{N}_p}{g_p}\right),
\end{equation}

\noindent where $V_{ij}$ denotes the mean electrostatic interaction energy between subshells $i$ and $j$, $\bar{N}_p$ the average electron population of subshell $n_p\ell_p$ and $g_p$ its degeneracy. Several years later, Honda \emph{et al.} \cite{KOIKE95,KOIKE97,HONDA97}  performed intensive MCDF (Multi-Configuration Dirac-Fock) calculations (using the GRASP2 code \cite{GRANT70,DYALL89}) and proposed to extrapolate the satellite emission spectra of gold ions from a single spectator to an arbitrary number of spectators. Such an approach is interesting, but inadequate when two or more spectator electrons are in the same shell. 

\begin{figure}
\vspace{10mm}
\begin{center}
\includegraphics[width=10cm]{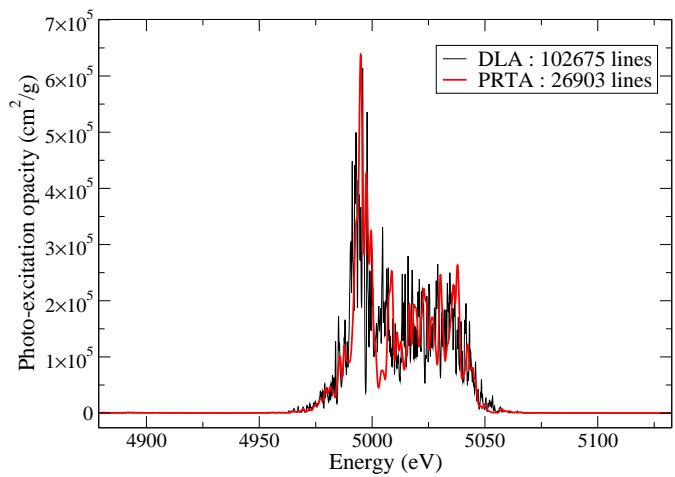}
\vspace{5mm}
\end{center}
\caption{(Color online) Comparison between two SCO-RCG calculations relying respectively on DLA and PRTA treatments of lines for transition array 3p$_{3/2}\rightarrow$ 5s in a Hg plasma at $T$=600 eV and $\rho$=0.01 g/cm$^3$. The DLA calculation contains 102 675 lines and the PRTA 26 903 lines.\label{Hg_prta3.1_bis}}
\end{figure}

More recently, Iglesias \emph{et al.} presented the PRTA (Partially Resolved Transition Array) model \cite{IGLESIAS12,IGLESIAS12b,IGLESIAS12c}, which consists in representing a transition array by a small-scale detailed calculation that preserves the known properties of the transition array (energy and variance \cite{BAUCHE79,BAUCHE82,BAUCHE85}) and yields improved higher-order moments. In the PRTA approach, open subshells are split in two groups. The main group includes the active electrons and the electrons that are strongly coupled to them. The other subshells are relegated to the secondary group. A small-scale DLA calculation is performed for the main group (omitting therefore the subshells in the secondary group) and a statistical approach for the secondary group assigns the missing variance to the lines, which can be obtained analytically \cite{BAUCHE82}. For instance, one may replace the transition array $C\rightarrow C'$

\begin{equation}
\left\{\begin{array}{l}
C=\mathrm{K^2L^8(3s)^2(3p)^2(3d)^3(4s)^1}\\
C'=\mathrm{K^2L^8(3s)^2(3p)^1(3d)^3(4s)^1(5s)^1}\nonumber
\end{array}
\right.
\end{equation}

\noindent which contains 102675 lines, by the ``reduced'' transition array $\tilde{C}\rightarrow\tilde{C}'$

\begin{equation}
\left\{\begin{array}{l}
\tilde{C}=\mathrm{K^2L^8(3s)^2(3p)^2(3d)^3}\\
\tilde{C}'=\mathrm{K^2L^8(3s)^2(3p)^1(3d)^3(5s)^1}\nonumber
\end{array}
\right.
\end{equation}

\noindent which has only 26903 lines. The missing variance, \emph{i.e.}, the contribution of 4s$^1$) is added as an additional broadening to the lines of $\tilde{C}\rightarrow\tilde{C}'$. The additivity of the contributions of active and passive subshells to the total variance is valid under the assumption that the radial integrals (Slater and spin-orbit) are the same for the initial $C$ and final $C'$ configurations. We can see for the example in Fig. \ref{Hg_prta3.1_bis} for the transition 3p$_{3/2}$ $\rightarrow$ 5s of Hg, that the simplified calculation is very close to the exact one, although the transition array is very asymmetrical.

\section{The SCO-RCG code}\label{scorcg}

When lines coalesce into broad unresolved patterns due to physical broadening mechanisms (Stark effect, electron collisions, autoionization, etc.), they can be handled by global methods \cite{BAUCHE88}. On the other hand, some transition arrays exhibit a small number of lines that must be taken into account individually. Those lines are important for plasma diagnostics, interpretation of spectroscopy experiments and for calculating the Rosseland mean, which is very sensitive to the hollows in the spectrum. For that reason, we have been developping over the past five years the ``hybrid'' opacity code SCO-RCG \cite{PORCHEROT11,PAIN13}, which combines fine-structure calculations with statistical methods, assuming local thermodynamic equilibrium. 

\subsection{Structure of the code}

Data required for the calculation of lines (Slater, spin-orbit and dipolar integrals) are provided by SCO (Super-configuration Code for Opacity) \cite{BLENSKI00}, which takes into account plasma screening and density effects on the wave-functions. Then, level energies and lines are calculated by an adapted routine (RCG) of Cowan's atomic structure code \cite{COWAN81} performing the diagonalization of the Hamiltonian matrix. Transition arrays for which a detailed treatment is not required or impossible are described statistically, by UTA (Unresolved Transition Arrays) \cite{BAUCHE79}, SOSA (Spin-Orbit Split Arrays) \cite{BAUCHE85} or STA (Super Transition Arrays) \cite{BAR89} formalisms implemented in SCO.

Some other approaches were proposed to overcome the limits of statistical methods. For instance, in order to improve the accuracy of the STA approach, Hazak and Kurzweil \cite{HAZAK12} developed an interesting method, named CRSTA (Configurationally Resolved Super Transition Arrays); however, its degree of refinement is limited to the configurations, the terms structure being included statistically through UTA widths and shifts \cite{KURZWEIL13}.  

\subsection{Probabilities of levels, configurations and super-configurations in the hybrid formalism}

The total opacity is the sum of photo-ionization, inverse Bremsstrahlung and Thomson scattering spectra calculated by SCO code and a photo-excitation spectrum arising from contributions of SCO and Cowan's codes in the form

\begin{equation}
\kappa\left(h\nu\right)=\frac{1}{4\pi\epsilon_0}\frac{\mathcal{N}}{A}\frac{\pi e^2h}{mc}\sum_{X\rightarrow X'}f_{X\rightarrow X'}\mathcal{P}_X\Psi_{X\rightarrow X'}(h\nu),
\end{equation}

\noindent where $h$ is Planck's constant, $\mathcal{N}$ the Avogadro number, $\epsilon_0$ the vacuum polarizability, $m$ the electron mass, $A$ the atomic number and $c$ the speed of light. $\mathcal{P}$ is a probability, $f$ an oscillator strength, $\Psi(h\nu)$ a profile and the sum $X\rightarrow X'$ runs over lines, UTAs, SOSAs or STAs of all ion charge states present in the plasma. Special care is taken to calculate the probability of $X$, which can be either a level $\alpha J$ ($\alpha$ is a set of quantum numbers that uniquely specify the level and the coupling scheme used to obtain the total angular momentum $J$), a configuration $C$ or a super-configuration $S$, because it can be the starting point for different transitions (DLA, UTA, SOSA or STA). In order to ensure the normalization of probabilities, we introduce three disjoint ensembles: $\mathcal{D}$ (detailed levels $\alpha J$), $\mathcal{C}$ (configurations $C$ too complex to be detailed) and $\mathcal{S}$ (super-configurations $S$ that do not reduce to ordinary configurations). The total partition function then reads

\begin{equation}
U_{\mathrm{tot}}=U\left(\mathcal{D}\right)+U\left(\mathcal{C}\right)+U\left(\mathcal{S}\right)\;\;\;\;\mathrm{with}\;\;\;\;\mathcal{D}\cap\mathcal{C}\cap\mathcal{S}=\emptyset,
\end{equation}

\noindent where each term is a trace over quantum states of the form Tr$\left[e^{-\beta\left(\hat{H}-\mu\hat{N}\right)}\right]$, $\hat{H}$ being the Hamiltonian, $\hat{N}$ the number operator, $\mu$ the chemical potential and $\beta=1/\left(k_BT\right)$. The Hamiltonian takes into account the kinetic energy and the bound-bound, bound-free and free-free electron-electron interaction energies. The probabilities of the different species of the $N$-electron ion are

\begin{equation}
\mathcal{P}_{\alpha J}=\frac{1}{U_{\mathrm{tot}}}\left(2J+1\right)e^{-\beta\left(E_{\alpha J}-\mu N\right)},
\end{equation}

\noindent for a level belonging to $\mathcal{D}$,

\begin{equation}
\mathcal{P}_C=\frac{1}{U_{\mathrm{tot}}}\sum_{\alpha J\in C}\left(2J+1\right)e^{-\beta\left(E_{\alpha J}-\mu N\right)},
\end{equation}

\noindent for a configuration that can be detailed, 

\begin{equation}\label{probs}
\mathcal{P}_C=\frac{1}{U_{\mathrm{tot}}}g_C~e^{-\beta\left(E_C-\mu N\right)}
\end{equation}

\noindent for a configuration that can not be detailed (\emph{i.e.} belonging to $\mathcal{C}$) and

\begin{equation}
\mathcal{P}_S=\frac{1}{U_{\mathrm{tot}}}\sum_{C\in S}g_C~e^{-\beta\left(E_C-\mu N\right)}
\end{equation}

\noindent for a super-configuration. In the case where the transition $C\rightarrow C'$ is a UTA that can be replaced by a PRTA (see Fig. \ref{Hg_prta3.1_bis}), its contribution to the opacity is modified according to

\begin{equation}
f_{C\rightarrow C'}~\mathcal{P}_C~\Psi_{C\rightarrow C'}(h\nu)\approx\sum_{\bar{\alpha}\bar{J}\rightarrow\bar{\alpha'}\bar{J'}}f_{\bar{\alpha}\bar{J}\rightarrow\bar{\alpha'}\bar{J'}}~\mathcal{P}_{\bar{\alpha}\bar{J}}~\Psi_{\bar{\alpha}\bar{J}\rightarrow\bar{\alpha'}\bar{J'}}(h\nu),
\end{equation}

\noindent where the sum runs over PRTA lines $\bar{\alpha}\bar{J}\rightarrow\bar{\alpha'}\bar{J'}$ between pseudo-levels of the reduced configurations, $f_{\bar{\alpha}\bar{J}\rightarrow\bar{\alpha'}\bar{J'}}$ is the corresponding oscillator strength and $\Psi_{\bar{\alpha}\bar{J}\rightarrow\bar{\alpha'}\bar{J'}}$ is the line profile augmented with the statistical width due to the other (non included) spectator subshells. The probability of the pseudo-level $\bar{\alpha}\bar{J}$ of configuration $\bar{C}$ reads

\begin{equation}
\mathcal{P}_{\bar{\alpha}\bar{J}}=\frac{\left(2\bar{J}+1\right)e^{-\beta\left(E_{\bar{\alpha}\bar{J}}-\mu N\right)}}{\sum_{\bar{\alpha}\bar{J}\in\bar{C}}\left(2\bar{J}+1\right)e^{-\beta\left(E_{\bar{\alpha}\bar{J}}-\mu N\right)}}\times\mathcal{P}_C
\end{equation}

\noindent with

\begin{equation}
\sum_{\bar{\alpha}\bar{J}\in\bar{C}}\mathcal{P}_{\bar{\alpha}\bar{J}}=\mathcal{P}_C,
\end{equation}

\noindent where $\mathcal{P}_C$ is the probability of the genuine configuration given in Eq. (\ref{probs}).

\subsection{Interpretation of experiments}

\begin{figure}
\vspace{10mm}
\begin{center}
\includegraphics[width=10cm]{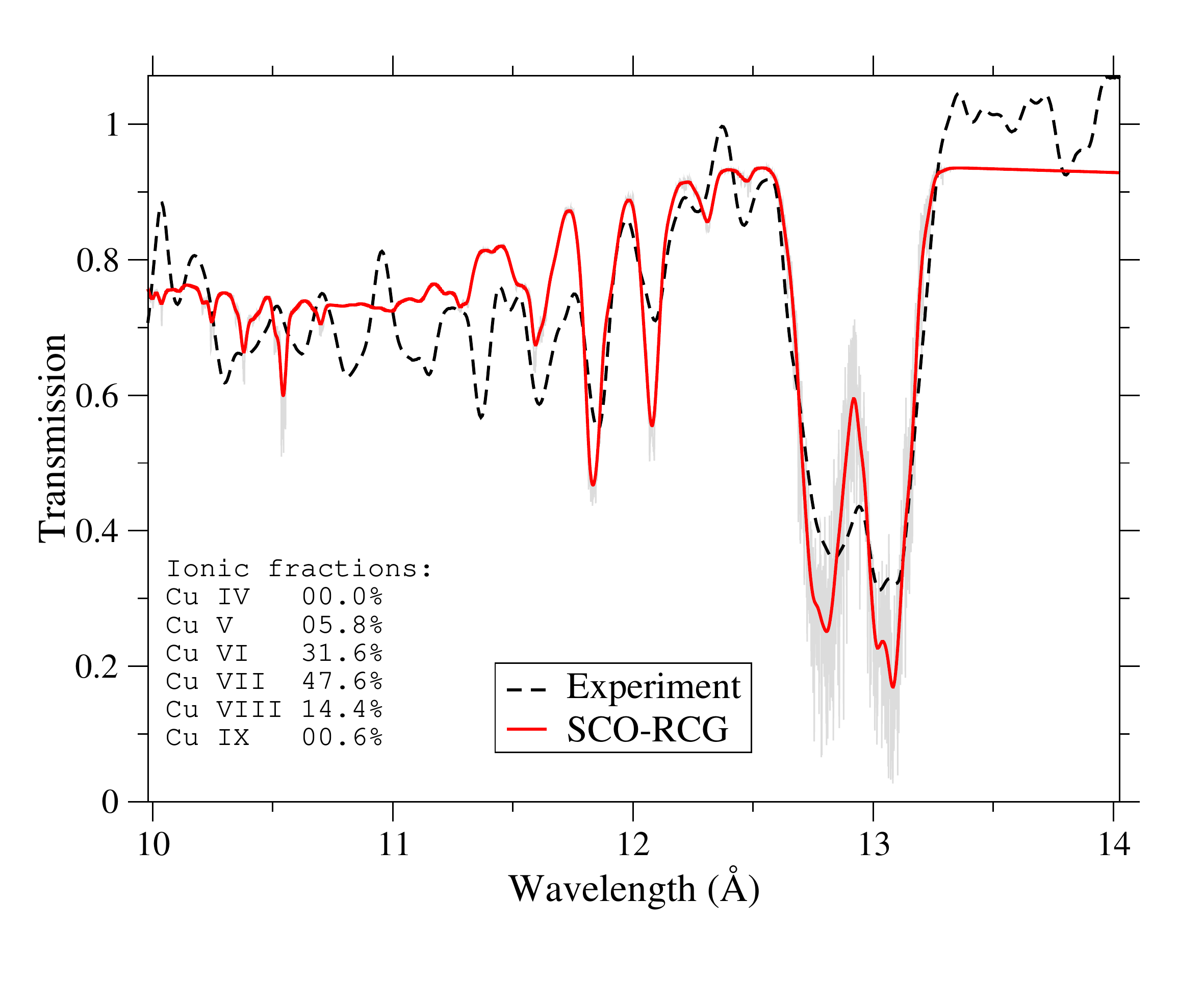}
\vspace{5mm}
\end{center}
\caption{(Color online) Interpretation with SCO-RCG code of the copper spectrum (2p$\rightarrow$ 3d transitions) measured by Loisel \emph{et al.} \cite{LOISEL09,BLENSKI11a,BLENSKI11b}. The temperature is $T$=16 eV, the density $\rho$=5 10$^{-3}$ g/cm$^3$, the areal mass $\rho L$=40 $\mu$g/cm$^2$ and the instrumental width 2.8 eV. The gray set of lines corresponds to the calculated spectrum before accounting for the intrumental broadening. \label{figure5}}
\end{figure}

\begin{figure}
\vspace{10mm}
\begin{center}
\includegraphics[width=10cm]{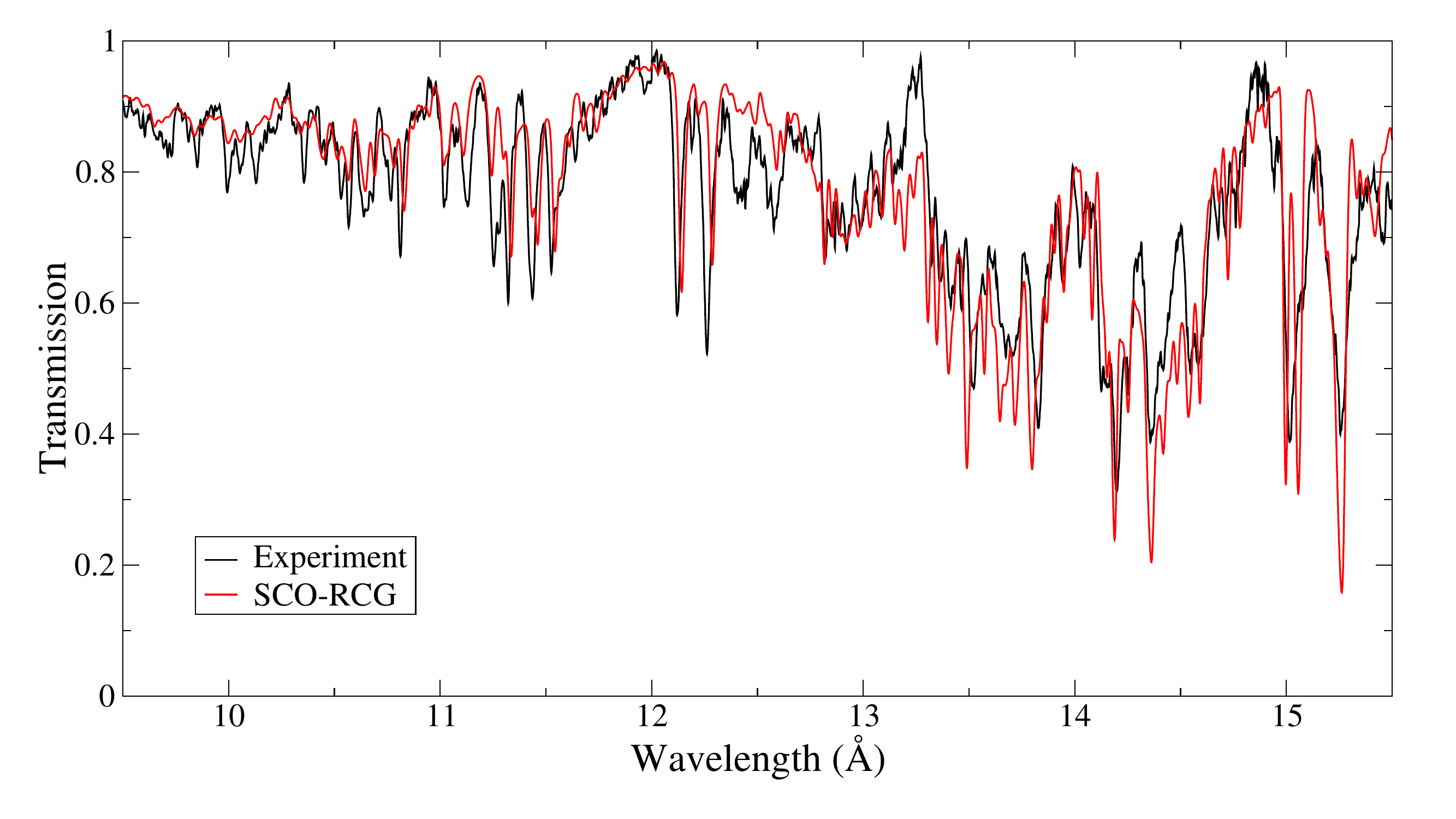}
\vspace{5mm}
\end{center}
\caption{(Color online) Interpretation with SCO-RCG code of the iron spectrum (2p$\rightarrow$ 3d transitions) measured by Bailey \emph{et al.} \cite{BAILEY07}. The temperature is $T$=150 eV, the density $\rho$=0.058 g/cm$^3$, the areal mass $\rho L$=32 $\mu$g/cm$^2$ and the instrumental width 1.8 eV.\label{fitBaileywithorwithoutSCO}}
\end{figure}

The SCO-RCG code has been successfully compared to spectra measured in experiments performed on laser-produced plasmas, for example, (Fig. \ref{figure5}) \cite{LOISEL09,BLENSKI11a,BLENSKI11b} and a Z-pinch, for example, (Fig. \ref{fitBaileywithorwithoutSCO}) \cite{BAILEY07} facility. As mentioned in Sec. \ref{sec1}, the quantity which is measured experimentally is the transmission, related to the opacity by Beer-Lambert-Bouguer's law: $T(h\nu)=e^{-\rho L \kappa(h\nu)}$, where $L$ is the thickness of the sample. This relation between transmission and opacity is valid under the assumption that the material is optically thin and that re-absorption processes can be neglected. 

In the present version of the code, configuration mixing is limited to electrostatic interaction between relativistic sub-configurations ($n\ell j$ orbitals) belonging to a non-relativistic configuration ($n\ell$ orbitals), namely ``relativistic configuration interaction''. That effect has a strong impact on the ratio of the two relativistic substructures of the 2p$\rightarrow$ 3d transition, visible around $\lambda$=13 \AA~in Fig. \ref{figure5}.

\subsection{Previous approach used in SCO-RCG}

\begin{figure}
\vspace{10mm}
\begin{center}
\includegraphics[width=8cm]{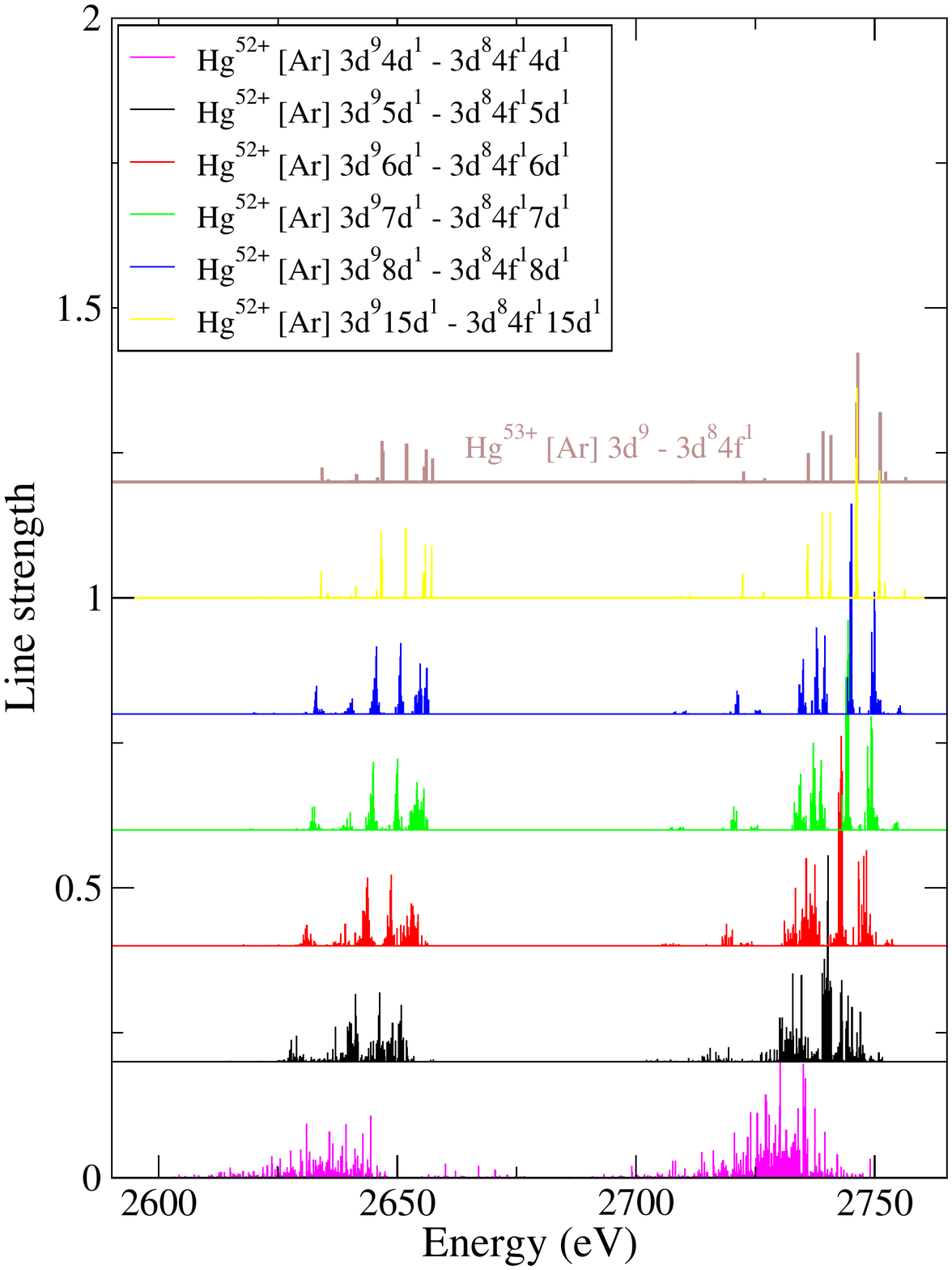}
\vspace{5mm}
\end{center}
\caption{(Color online) Effect of passive subshell $n$d$^1$, $n$=4,5,6,7,8 and 15, on transition array [Ar] 3d$^9$ $\rightarrow$ 3d$^8$4f$^1$. The calculation was performed with Cowan's code (RCN/RCN2/RCG).\label{Hg_specnd}}
\end{figure}

\begin{figure}
\vspace{10mm}
\begin{center}
\includegraphics[width=8cm]{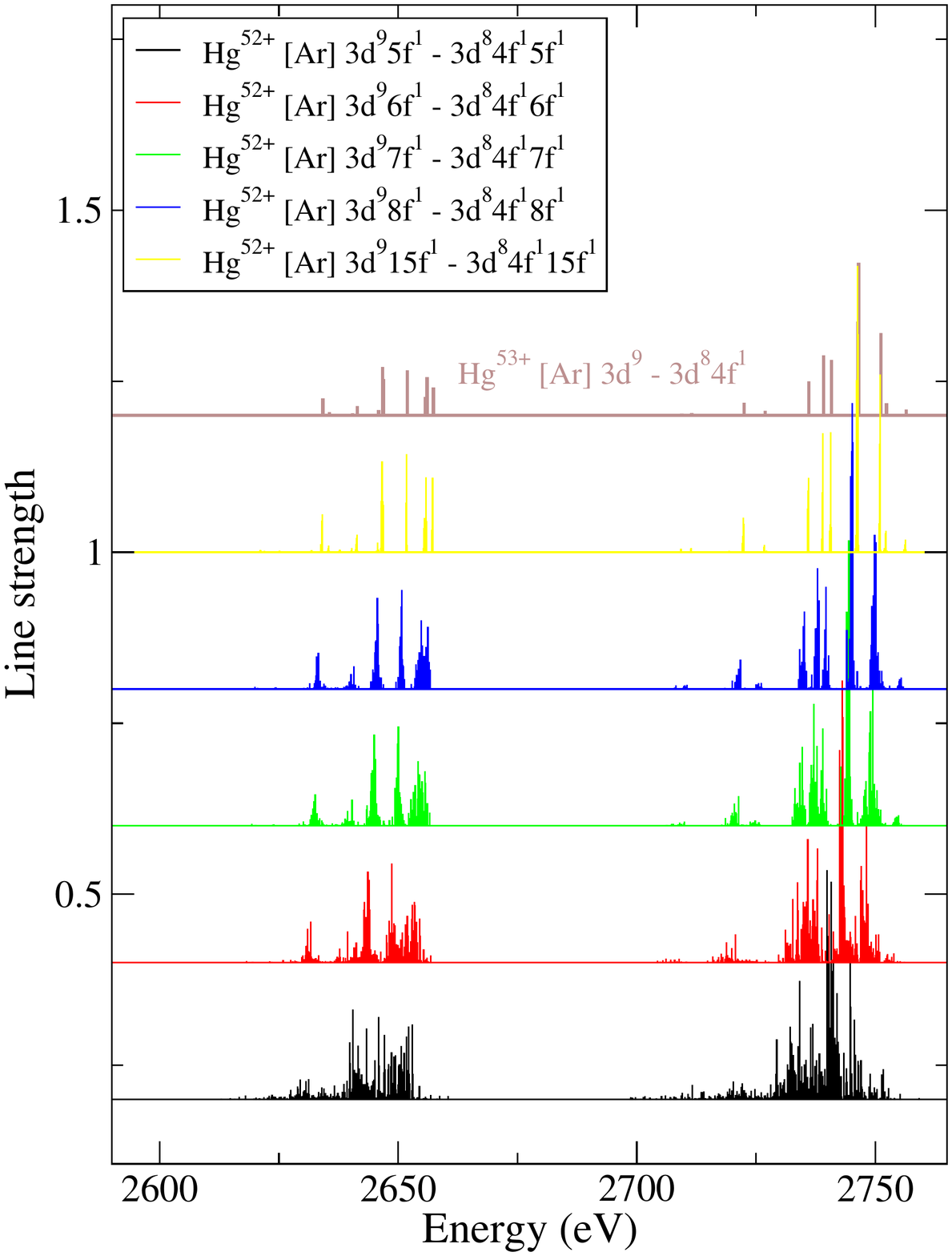}
\vspace{5mm}
\end{center}
\caption{(Color online) Effect of passive subshell $n$f$^1$, $n$=4,5,6,7,8 and 15, on transition array [Ar] 3d$^9$ $\rightarrow$ 3d$^8$4f$^1$. The calculation was performed with Cowan's code (RCN/RCN2/RCG).\label{Hg_specnf}}
\end{figure}

Figures \ref{Hg_specnd} and \ref{Hg_specnf} illustrate the effect of adding a passive subshell $n$d$^1$ and $n$f$^1$ respectively, with $n$=4,5,6,7,8 and 15, to the transition array 3d$^9$ $\rightarrow$ 3d$^8$4f$^1$. The calculation was performed with Cowan's suite of codes (RCN/RCN2/RCG). We can see that when the spectator electron is 4d, the spectrum of 3d$^9$4d$^1$$\rightarrow$ 3d$^8$4d$^1$4f$^1$ is very different from the one of 3d$^9$$\rightarrow$ 3d$^8$4f$^1$. This is due to the fact that the spectator electron 4d is strongly coupled to the active ones 3d and 4f. When the electron is pushed away (up to 15d and 15f), the spectrum becomes closer and closer to the resonance lines.

In SCO-RCG \cite{PORCHEROT11,PAIN13}, orbitals are treated individually up to a certain limit beyond which all remaining orbitals (up to 15t) are gathered in a single super-shell $\sigma$, called attributively ``Rydberg super-shell'' in the following. The grouped orbitals are chosen so that they weakly interact with inner orbitals. The last orbital of the super-shell corresponds to the one with the highest $n$ and $\ell$ values for which a wavefunction can be calculated within the ion-sphere model. This corresponds to the pressure-ionization limit beyond which orbitals are assumed to be in the continuum. The grouped orbitals can also be defined by the user. In the latter case, it is useful to have an estimate of the last populated shell. For high values of $n$, the members of the series are found to be closer and closer together. When the finite width of the lines is also taken into account, then, for the highest levels, the spacing between two neighboring lines can be smaller than the line's widths. Beyond this limit the lines cannot be resolved as separate lines any longer, but rather generate a smooth continuum. For the hydrogen-like Lyman series, the highest line that is still resolvable was calculated by Inglis and Teller  \cite{INGLIS39} assuming a Stark broadening. The expression of the maximal principal quantum number $n_{\mathrm{max}}$, determined by Inglis and Teller from a simplified screening model in a very dilute hydrogen plasma, was extended to non-hydrogenic system for instance by Alaterre \cite{ALATERRE84}:

\begin{equation}\label{it}
\log_{10} N_e=23.26+4\log_{10} Z^*-7.5\log_{10} n_{\mathrm{max}}.
\end{equation}

\noindent which gives

\begin{equation}\label{it2}
n_{\mathrm{max}}\approx\left[\frac{1.8197\;10^{23}}{N_e\left[cm^{-3}\right]}Z^{*4}\right]^{\frac{2}{15}}.
\end{equation}

\noindent This formula is slightly different from the one proposed by Griem \cite{GRIEM97} ($4\log_{10} Z^*$ instead of $4.5\log_{10} Z^*$ in Eq. (\ref{it})), who considered the electronic microfield instead of the ionic microfield for Eq. (\ref{it}).

If the population $Q$ of the Rydberg super-shell $\sigma$ is equal to zero, the opacity calculation involves only configurations whose contribution can be potentially detailed and therefore consists only of UTAs, SOSAs or DLAs. Indeed, a one-electron jump from an orbital $\left(n\ell\right)$ gives rise to a transition array which can be detailed or not. In order to decide whether a line-by-line calculation is necessary and to determine the validity of statistical methods, it can be interesting to quantify the porosity (localized absence of lines) of transition arrays \cite{GAFFNEY11a,GAFFNEY11b}. In SCO-RCG, the most limiting parameter is the maximum number of lines per transition array allowed by the user. There are also some computational constraints, mostly due to the sizes of arrays in the RCG routine: the number of lines must not exceed 800,000, the size of a $J$-block (number of levels having the same value of total atomic angular momentum $J$) of the Hamiltonian must not exceed 4,000 and we can not handle configurations with more than two electrons (or more than two vacancies) in any orbital with $\ell\ge$ 4 (due to the tabulated coefficients of fractional parentage \cite{NIELSON63,KARAZIJA68}). Optionally, to speed up calculations, it is also possible to resort to a criterion in order to check the coalescence of the transition array. For instance, we may define the parameter

\begin{equation}
\xi=\frac{\bar{d}}{w_{\mathrm{ph}}}
\end{equation}

\noindent where $w_{\mathrm{ph}}$ is the largest physical broadening (natural, Doppler, Stark, collisions, autoionization, etc.). In SCO-RCG, all the lines in a given transition array have the same physical broadenings and the average distance (in energy) $\bar{d}$ between two lines in the transition arrays is evaluated as

\begin{equation}
\bar{d}=\frac{\sqrt{v_{\mathrm{UTA}}}}{L},
\end{equation}

\noindent $v_{\mathrm{UTA}}$ being the UTA variance \cite{BAUCHE79} and $L$ the number of lines, which can be evaluated exactly by efficient recursive techniques or estimated by an approximate formula \cite{GILLERON09}. If $\xi$ is smaller than a particular value $\epsilon$ with $\epsilon<<1$, the contribution of the transition array to opacity is calculated statistically within the UTA or SOSA formalisms. 

In one option of the code, if $Q\ge 1$, any transition starting from this super-configuration is treated in the STA formalism. However, such statistical modeling of Rydberg spectators may be irrelevant for some well-resolved transition arrays, as we will see in the next section.

\section{Test case: F-like 2p$\rightarrow$ 4d transitions of iron. Is the statistical modeling of Rydberg spectators relevant?}\label{test}

\begin{figure}
\vspace{10mm}
\begin{center}
\includegraphics[width=10cm]{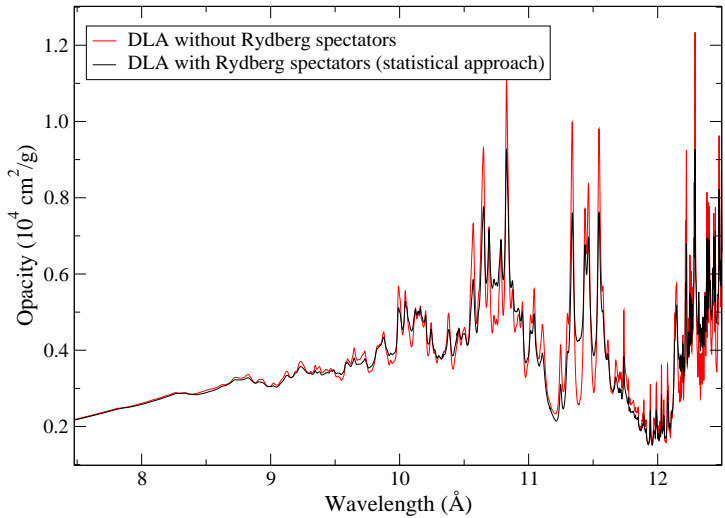}
\vspace{5mm}
\end{center}
\caption{(Color online) DLA computation of the opacity of iron at $T$=182 eV and $N_e$=3.1~10$^{22}$ cm$^{-3}$ without spectators and with spectators treated in a statistical way as a Rydberg super-shell (5s $\cdots$ 8d)$^1$.\label{figure_loisel2014_ter}}
\end{figure}

\begin{figure}
\vspace{10mm}
\begin{center}
\includegraphics[width=10cm]{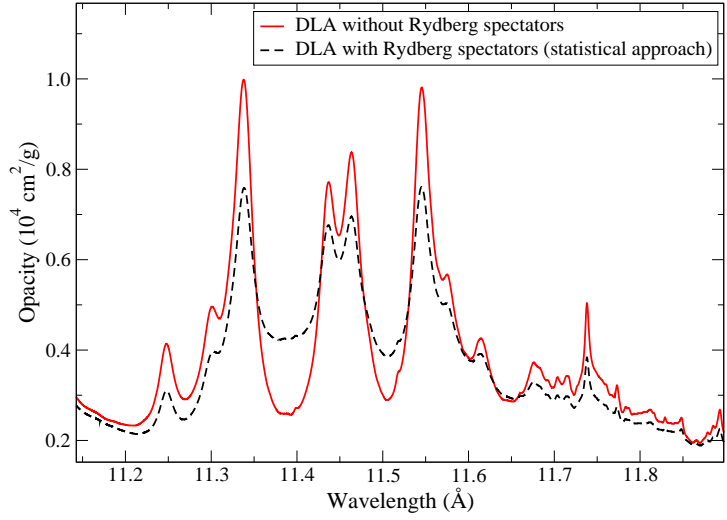}
\vspace{5mm}
\end{center}
\caption{(Color online) Detail of Fig. \ref{figure_loisel2014_ter} in the range of transitions F-like [Be] 2p$^5\rightarrow$ 2p$^4$ 4d$^1$.\label{figure_loisel2014_zoom_ter}}
\end{figure}

In the case of an iron plasma at $T$=182 eV and $N_e$=3.1~10$^{22}$ cm$^{-3}$ (conditions of a recent experiment performed by J. Bailey \cite{BAILEY15})\footnote{In the present paper, we will not show any experimental spectrum nor discuss any detailed aspects of the experiment (for further explanations see Ref. \cite{BAILEY15}).}, the statistical modeling of satellites [Be] 2p$^5$(5s $\cdots$ 8d)$^1\rightarrow$ 2p$^4$4d$^1$(5s $\cdots$ 8d)$^1$ fill significantly the gaps around 11.4 and 11.5 \AA~(see Figs. \ref{figure_loisel2014_ter} and \ref{figure_loisel2014_zoom_ter}). To check whether it is relevant or not, we have studied the effect of each spectator electron separately (from 5s to 8d) on the spectrum.

\subsection{Study of the impact of spectators individually}

\begin{figure}
\vspace{10mm}
\begin{center}
\includegraphics[width=10cm]{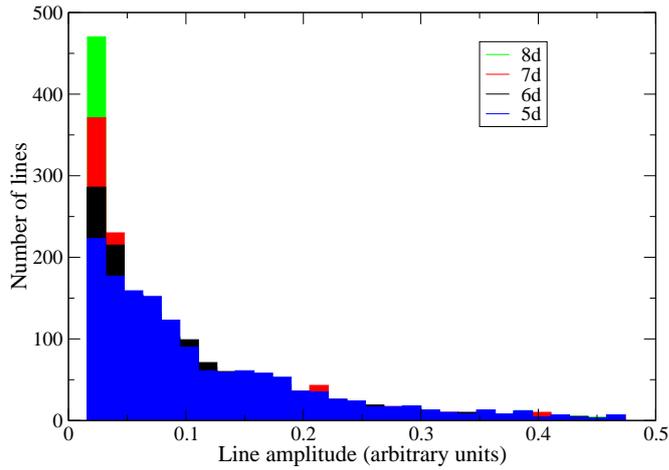}
\vspace{5mm}
\end{center}
\caption{(Color online) Histogram of the amplitude (square root of the line strength) for 2p$^5n$d$^1\rightarrow$ 2p$^4$4d$^1n$d$^1$ with $n$=5 to 8. The calculation was performed with Cowan's code (RCN/RCN2/RCG).\label{histogram}}
\end{figure}

\begin{figure}
\vspace{10mm}
\begin{center}
\includegraphics[width=8cm]{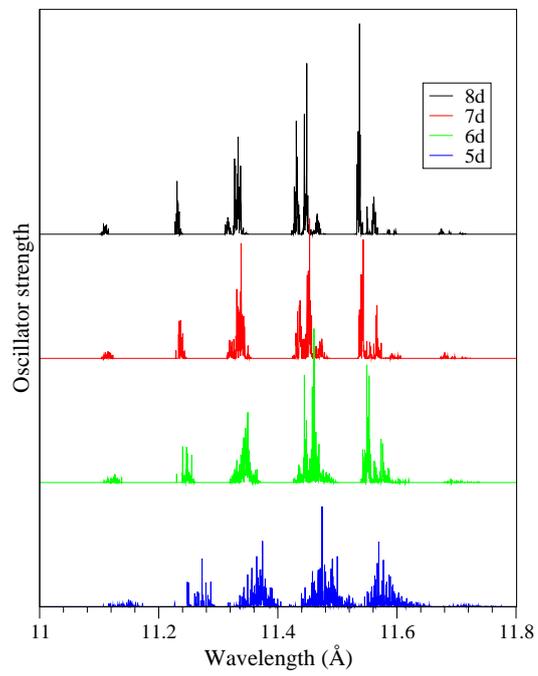}
\vspace{15mm}
\end{center}
\caption{(Color online) Oscillator strength of 2p$^5\rightarrow$ 2p$^4$ 4d$^1$ with a spectator electron $n$d$^1$, $n$=5 to 8.\label{figure2_bis_fseul}}
\end{figure}

\begin{figure}
\vspace{10mm}
\begin{center}
\includegraphics[width=10cm]{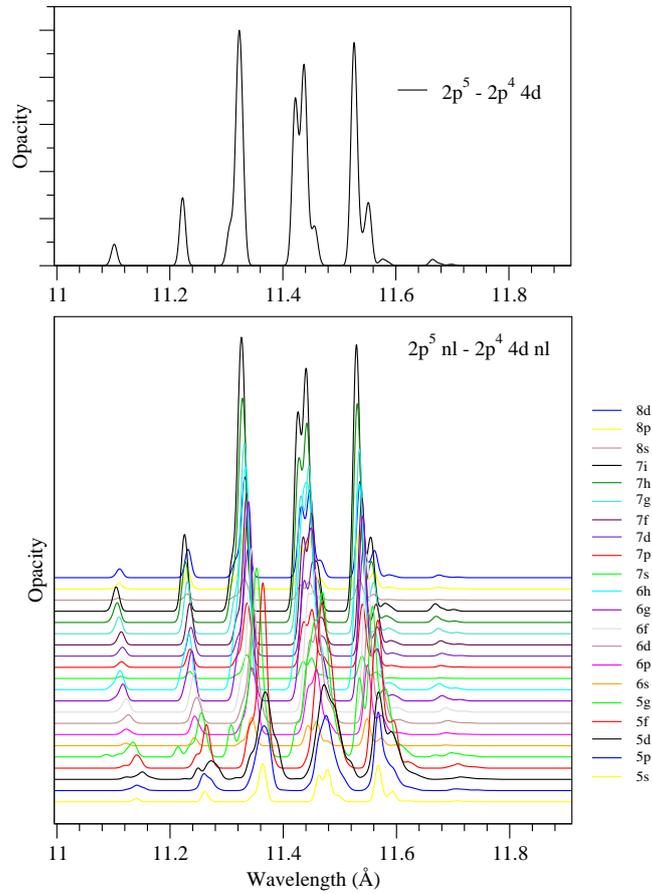}
\vspace{5mm}
\end{center}
\caption{(Color online) Opacity of transition array 2p$^5\rightarrow$ 2p$^4$ 4d$^1$ without any spectator and with a spectator electron $n\ell^1$ (from 5s to 8d). An instrumental resolution of 0.005 \AA~ was chosen for clarity reasons.\label{figure2bis}}
\end{figure}

\begin{figure}
\vspace{10mm}
\begin{center}
\includegraphics[width=10cm]{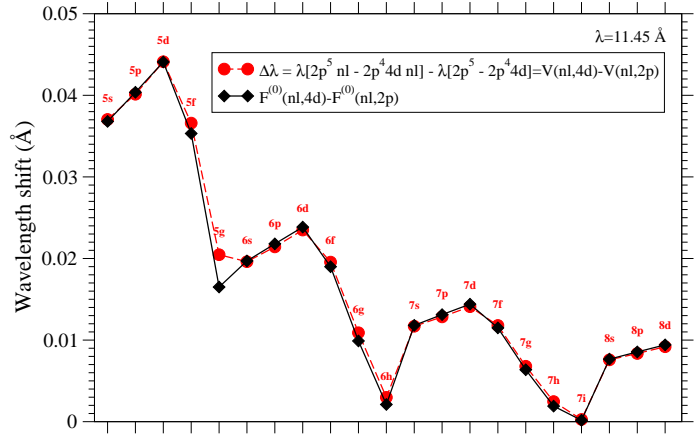}
\vspace{10mm}
\end{center}
\caption{(Color online) Shift of the center of gravity (first-order strength-weighted moment) of transition array 2p$^5\rightarrow$ 2p$^4$ 4d$^1$ due to a spectator electron $n\ell^1$ (from 5s to 8d) compared to the difference $F^{(0)}\left(n\ell,4d\right)-F^{(0)}\left(n\ell,2p\right)$. The calculations were performed with Cowan's code (RCN/RCN2/RCG) and checked with formula (\ref{inte}).\label{figure3_ter}}
\end{figure}

\begin{figure}
\vspace{10mm}
\begin{center}
\includegraphics[width=10cm]{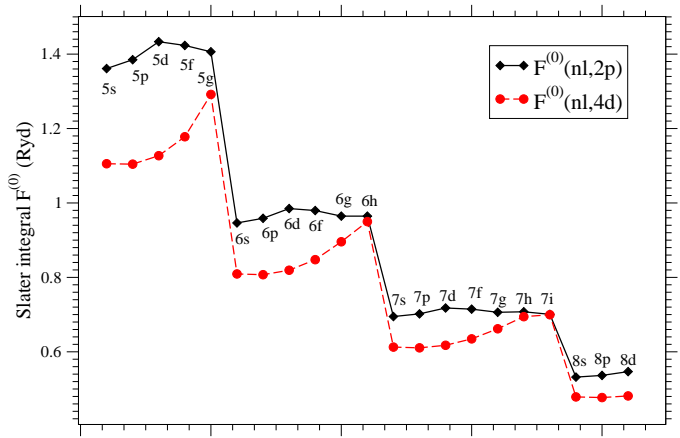}
\vspace{10mm}
\end{center}
\caption{(Color online) Comparison of Slater integrals $F^{(0)}(n\ell,2p)$ and $F^{(0)}(n\ell,4d)$ for $n\ell$ varying from 5s to 8d. The calculation was performed with Cowan's code (RCN/RCN2/RCG).\label{fig_supp_fg1}}
\end{figure}

When the spectator electron moves from an s to an f orbital, the number of lines increases from 860 to 1724 (see table \ref{tab:a}). The histogram of the amplitudes (square root of the line strength) of the lines  for 2p$^5n$d$^1\rightarrow$ 2p$^4$4d$^1n$d$^1$ with $n$=5 to 8 (see Fig. \ref{histogram}) shows that the number of weak lines increases when the spectator moves far away from the nucleus (see table \ref{tab:a1}). The detailed calculation of satellites differs then from resonance lines only by a small shift and broadening (see Figs. \ref{figure2_bis_fseul} and \ref{figure2bis}). The shift of the center of gravity of transition array 2p$^5\rightarrow$ 2p$^4$4d$^1$ due to the adding of a spectator electron is shown in Fig. \ref{figure3_ter}. It decreases globally as a function of $n$ with a non-monotonic behaviour inside a given shell: a slight increase for s$\rightarrow$p$\rightarrow$d and then a decrease for $\ell\ge$3. We can see (solid line of Fig. \ref{figure3_ter}) that the main contribution to the shift stems from the direct $F^{(0)}$ Slater integrals, for which the respective variations are displayed in Fig. \ref{fig_supp_fg1}. It is worth noting that the quantity $F^{(0)}(n\ell,2p)$ increases for $\ell=0,1,2$ and decreases for $\ell\ge$3 inside a particular shell, while $F^{(0)}(n\ell,4d)$ is an increasing function of $\ell$ inside a shell.

\begin{table*}[t]
\begin{center}
\begin{tabular}{|c|c|}\hline
Spectator $n\ell$ & Number of lines \\ \hline
p & 860 \\ \hline
d & 1506 \\\hline
f & 1707 \\\hline
g & 1724 \\\hline
\end{tabular}
\caption{Number of lines of transition array 2p$^5n\ell\rightarrow$ 2p$^4$ 4d$^1n\ell^1$, $\ell$ varying from 1 to 4.}\label{tab:a}
\end{center}
\end{table*}

\begin{table*}[t]
\begin{center}
\begin{tabular}{|c|c|}\hline
Spectator $\ell$ & Number of lines with $f<f_{\mathrm{max}}/1000$ \\ \hline
5p & 201 \\\hline
6p & 254 \\\hline
7p & 286 \\\hline
8p & 310 \\\hline\hline
5d & 342 \\\hline
6d & 468 \\\hline
7d & 569 \\\hline
8d & 652 \\\hline\hline
5f & 530 \\\hline
6f & 641 \\\hline
7f & 745 \\\hline
8f & 820 \\\hline\hline
\end{tabular}
\caption{Number of lines of transition array 2p$^5n\ell\rightarrow$ 2p$^4$ 4d$^1n\ell^1$ for which the oscillator strength $f$ is smaller than its maximum value divided by 1000.}\label{tab:a1}
\end{center}
\end{table*}

\subsection{Failure of the statistical modeling of Rydberg spectators\label{proce}}

\begin{figure}
\vspace{10mm}
\begin{center}
\includegraphics[width=10cm]{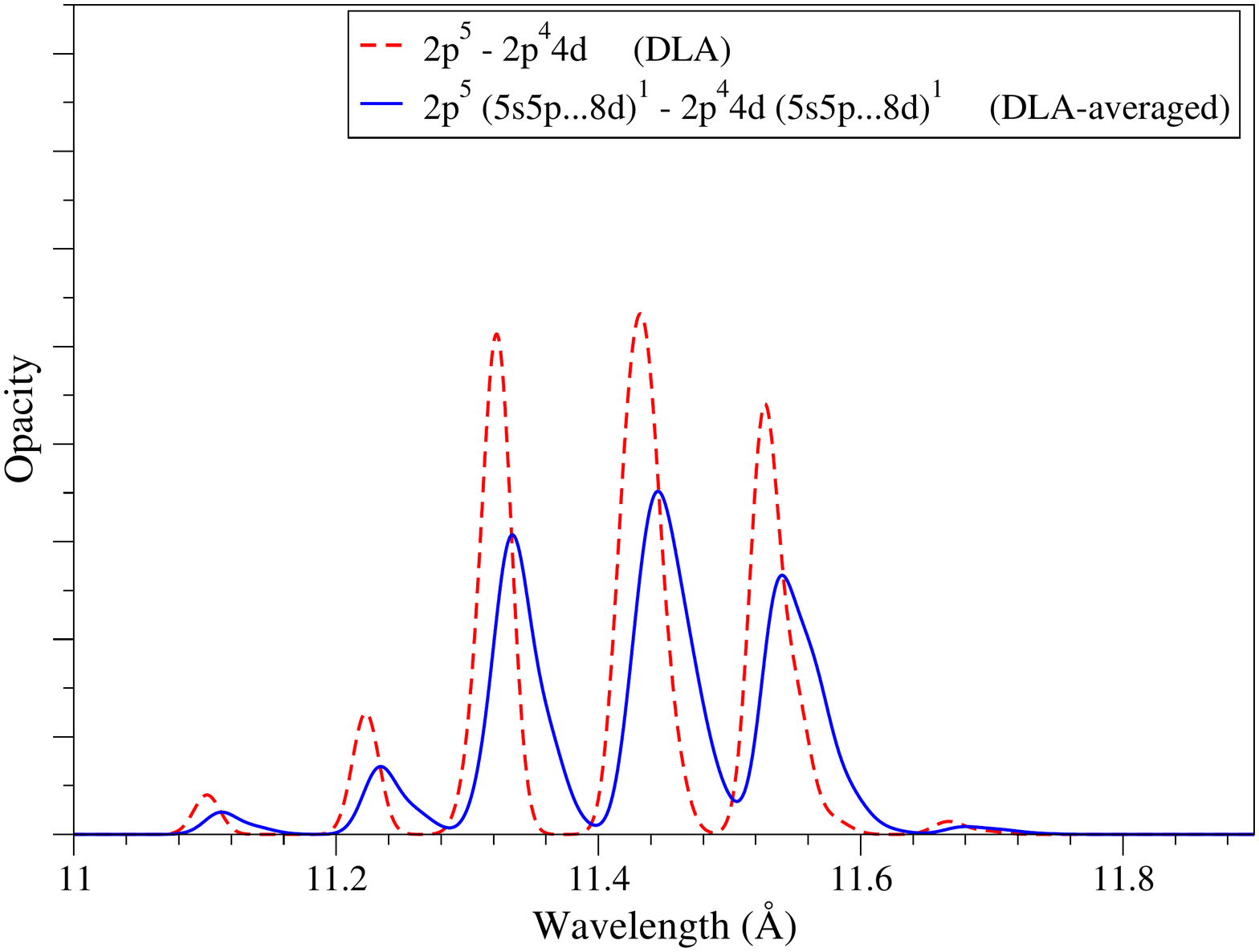}
\vspace{5mm}
\end{center}
\caption{(Color online) Opacity of transition array 2p$^5\rightarrow$ 2p$^4$ 4d$^1$ without any spectator (DLA) and with a spectator electron $n\ell^1$ (from 5s to 8d). ``DLA-averaged'' means that the spectrum was averaged over the different spectra corresponding to a different spectator electron, as described in Sec. \ref{proce}. The calculations were carried out with Cowan's suite of codes (RCN/RCN2/RCG) and an instrumental resolution of 0.01 \AA~ was chosen for clarity reasons.\label{figure2quad}}
\end{figure}

We have first calculated the contribution to the total opacity of F-like lines of 1s$^2$2s$^2$2p$^5\rightarrow$ 1s$^2$2s$^2$2p$^4$4d$^1$ (dashed line of Fig. \ref{figure2quad}). We have then calculated the contribution arising from Ne-like lines of the type 1s$^2$2s$^2$2p$^5n\ell\rightarrow$ 1s$^2$2s$^2$2p$^4$4d$^1n\ell$ with $n\ell$ ranging from 5s to the last bound orbital 8d.

In order to investigate the contribution of satellites, we proceed as follows. Considering $C_i=C_0\left(n_i\ell_i\right)^1$ with $C_0$=1s$^2$2s$^2$2p$^5$ and $n_i\ell_i$ ranging from 5s to 8d, we compare, using Cowan's code (RCN/RCN2/RCG), the pure DLA contribution to opacity (\emph{i.e.} neglecting the contribution of the Rydberg electrons) of $C_0\rightarrow C'_0$=1s$^2$2s$^2$2p$^4$4d$^1$, evaluated as

\begin{equation}\label{pureDLA}
\kappa(h\nu)=\frac{1}{4\pi\epsilon_0}\frac{\mathcal{N}}{A}\frac{\pi e^2h}{mc}\frac{1}{\mathcal{Z}}\sum_{\alpha J\in C_0}g_{\alpha J}e^{-\beta\left(E_{\alpha J}-\mu Q_0\right)}f_{\alpha J\rightarrow\alpha'J'}\psi_{\alpha J\rightarrow\alpha'J'}(h\nu)
\end{equation}

\noindent with $Q_0$=9, to the ``DLA-averaged'' opacity of $C_i\rightarrow C'_i$=1s$^2$2s$^2$2p$^4$4d$\left(n_i\ell_i\right)^1$, calculated as

\begin{equation}\label{DLAaveraged}
\kappa(h\nu)=\frac{1}{4\pi\epsilon_0}\frac{\mathcal{N}}{A}\frac{\pi e^2h}{mc}\frac{1}{\mathcal{Z}}\sum_{C\in\left\{C_1,\cdots,C_N\right\}}\sum_{\alpha J\in C}g_{\alpha J}e^{-\beta\left(E_{\alpha J}-\mu Q\right)}f_{\alpha J\rightarrow\alpha'J'}\psi_{\alpha J\rightarrow\alpha'J'}(h\nu)
\end{equation}

\noindent with $Q=10$ and $n_i\ell_i$ ranging from 5s to 8d. $\mathcal{Z}$ is the partition function. As can be seen in Fig. \ref{figure2quad}, the total contribution of satellites (solid line), calculated by Eq. (\ref{pureDLA}), is very similar to the resonance transition (dashed curve) calculated by Eq. (\ref{DLAaveraged}), except that it is slightly shifted and broadened. These spectator electrons are very weakly bound to the ion, so that they do not interact much with the core electrons and perturb the transition very weakly. Therefore, the ``DLA-averaged'' spectrum is very close to the resonance spectrum, \emph{i.e.}, the pure DLA calculation without any spectators.

We then repeated the calculation by using the statistical approach to model the effect of the spectator. All orbitals from 5s to 8d are now gathered in a single super-shell and the resulting transition array 1s$^2$2s$^2$2p$^5\left(\right.$5s $\cdots$ 8d$\left.\right)^1\rightarrow$ 1s$^2$2s$^2$2p$^4$4d$^1\left(\right.$5s $\cdots$ 8d$\left.\right)^1$ (see Fig. \ref{fig1}) is now represented by only two unresolved relativistic substructures 2p$_{1/2}\rightarrow$ 4d$_{3/2}$, 2p$_{3/2}\rightarrow$ 4d$_{5/2}$, while 2p$_{3/2}\rightarrow$ 4d$_{3/2}$ is not visible. Therefore, it can not reproduce the porosity of the transition arrays. Moreover and unfortunately, in that case, the maxima of the statistical spectral features correspond to minima of the detailed spectrum. In other words, the statistical calculation fills the gap between the lines more efficiently than the detailed one.

\section{New approach}\label{new}

Following the PRTA approach \cite{IGLESIAS12,IGLESIAS12b,IGLESIAS12c} mentioned in the introduction, we suggest the replacement of the super-transition array $S\rightarrow S'$

\begin{equation}
\left\{\begin{array}{l}
S=\mathrm{K^2L^8(3s)^2(3p)^6(3d)^{10}(4s)^2(4p)^3(4d)^4(4f)^6(5s5p \cdots 8s8p8d)^Q}\\
S'=\mathrm{K^2L^8(3s)^2(3p)^6(3d)^9(4s)^2(4p)^3(4d)^4(4f)^7(5s5p\cdots 8s8p8d)^Q}\nonumber
\end{array}
\right.
\end{equation}

\noindent by the ``reduced'' transition array $\tilde{C}\rightarrow\tilde{C}'$

\begin{equation}
\left\{\begin{array}{l}
\tilde{C}=\mathrm{K^2L^8(3s)^2(3p)^6(3d)^{10}(4s)^2(4p)^3(4d)^4(4f)^6}\\
\tilde{C}'=\mathrm{K^2L^8(3s)^2(3p)^6(3d)^9(4s)^2(4p)^3(4d)^4(4f)^7}\nonumber
\end{array}
\right.
\end{equation}

\noindent and under the assumption that the Slater integrals are the same for the initial and final (super-)configurations $S$ and $S'$, the variance is additive with respect to the separation between active and passive subshells:

\begin{equation}
v_{S\rightarrow S'}^{\mathrm{lines}}=v_{\tilde{C}\rightarrow \tilde{C}'}^{\mathrm{lines}}+v_{\mathrm{spect}}\left[\mathrm{(5s5p5d5f5g\cdots8s8p8d)^Q}\right].
\end{equation}

\noindent Therefore, the effect of spectators from 5s to 8d is included as an additional Gaussian broadening of variance $v_{\mathrm{spect}}\left[\mathrm{(5s5p5d5f5g\cdots8s8p8d)^Q}\right]$ of the detailed lines of $\tilde{C}\rightarrow \tilde{C}'$.

\subsection{Contribution of a spectator electron $\ell_p^{N_p}$ to the statistical shift and variance of transition array $\ell_{\alpha}^{N+1}\ell_{\beta}'^{N'}\rightarrow\ell_{\alpha}^N\ell_{\beta}'^{N'+1}$}

The energy of a configuration $C$ reads

\begin{equation}
E_C=\sum_sN_sI_s+\frac{1}{2}\sum_{s,s'}N_s\left(N_{s'}-\delta_{s,s'}\right)V_{ss'}
\end{equation}

\noindent where $N_s$ is the electron population of subshell $s$ and $\delta_{s,s'}$ the Kronecker symbol. The one-body integrals are defined as

\begin{equation}
I_s=\epsilon_s+\int_0^{\infty}\left[eV(r)-\frac{Ze^2}{4\pi\epsilon_0r}\right]y_s^2(r)dr
\end{equation}

\noindent where $y_s$, radial part of the wavefunction multiplied by $r$, obeys Schr\"odinger equation

\begin{equation}
\left[-\frac{\hbar^2}{2m}\frac{d^2}{dr^2}+\frac{\hbar^2\ell(\ell+1)}{2mr^2}-eV(r)\right]y_s(r)=\epsilon_sy_s(r),
\end{equation}

\noindent and the two-body interactions are given by

\begin{equation}\label{inte}
V_{ij}=\frac{g_i}{g_i-\delta_{ij}}\left[R^{(0)}(ij,ij)-\frac{1}{2}\sum_k\threej{\ell_i}{k}{\ell_j}{0}{0}{0}^2R^{(k)}(ij,ji)\right],
\end{equation}

\noindent $R^{(k)}$ being the configuration-interaction radial integral of order $k$ ($R^{(k)}(ij,ij)=F^{(k)}(ij)$ represents the direct Slater integral and $R^{(k)}(ij,ji)=G^{(k)}(ij)$ the exchange Slater integral). In the general case, the statistical shift due to $N_p$ spectators in subshell $\ell_p$  reads

\begin{eqnarray}
& &N_p\left(I_p^{C'}-I_p^{C}+V_{p\beta}^{C'}-V_{p\alpha}^C+\sum_{k\ne p}N_k\left(V_{pk}^{C'}-V_{pk}^C\right)\right)\nonumber\\
& &+\frac{1}{2}N_p\left(N_p-1\right)\left[V_{pp}^{C'}-V_{pp}^C\right].
\end{eqnarray}

\noindent When Slater integrals are assumed to be the same for $C$ and $C'$, the shift becomes 

\begin{equation}\label{shift}
N_p\left(V_{p\beta}-V_{p\alpha}\right), 
\end{equation}

\noindent where

\begin{eqnarray}
V_{p\beta}-V_{p\alpha}&=&F^{(0)}\left(n_{\beta}\ell_{\beta},n_p\ell_p\right)-F^{(0)}\left(n_{\alpha}\ell_{\alpha},n_p\ell_p\right)\nonumber\\
& &-\frac{1}{2}\sum_k\left\{\threej{\ell_{\beta}}{k}{\ell_p}{0}{0}{0}^2G^{(k)}\left(n_{\beta}\ell_{\beta},n_p\ell_p\right)\right.\nonumber\\
& &-\left.\threej{\ell_{\alpha}}{k}{\ell_p}{0}{0}{0}^2G^{(k)}\left(n_{\alpha}\ell_{\alpha},n_p\ell_p\right)\right\}.
\end{eqnarray}

\noindent In practice, in SCO-RCG, this shift is not required, since each super-configuration has its own self-consistent potential and its own set of wavefunctions.

For the contribution of spectator subshell $\ell_p^{N_p}$ to the variance, the general formula is more complicated, and can be found in Ref. \cite{KARAZIJA88}. When Slater integrals are assumed to be the same for $C$ and $C'$, the formula reduces to a more compact form and reads

\begin{equation}
\frac{N_p\left(4\ell_p+2-N_p\right)}{\left(4\ell_p+1\right)}\Delta_p^{(\alpha\beta)},
\end{equation}

\noindent where the two-electron variance $\Delta_p^{(\alpha\beta)}=\Delta\left(\ell_p\ell_{\alpha}\rightarrow\ell_p\ell_{\beta}\right)$ is given in table II of Ref. \cite{BAUCHE79} (formulas E1 to E8 and E'7, p. 2430).

\subsection{Contribution of a spectator super-shell $\sigma^Q$ to the statistical shift and variance}

Let us consider a super-shell $\sigma^Q$ (for instance, in the preceding case, $\sigma=(\mathrm{5s\:5p\:5d\:5f\:5g\cdots8s\:8p\:8d})$. If neither $\alpha$ nor $\beta$ do belong to $\sigma$, then the contribution of the ``spectator'' super-shell $\sigma$ to the center of gravity reads

\begin{equation}
\delta E^{\mathrm{spect}}_{\alpha\beta}[Q]=\sum_{n=1}^Q\frac{U_{Q-n}}{U_Q}\phi_n(D)
\end{equation}

\noindent with

\begin{equation}\label{phim}
\phi_m(D)=-\sum_{s\in\sigma}\left(-X_s\right)^mg_sD_s^{\alpha\beta},
\end{equation}

\noindent where $X_s=e^{-\beta\left(\epsilon_s-\mu\right)}$, $D_0^{\alpha\beta}=I_{\beta}-I_{\alpha}$ and $D_s^{\alpha\beta}=V_{s\beta}-V_{s\alpha}$. The usual STA partition functions $U_Q$ can be calculated by recursion relations \cite{GILLERON04}. The variance to be included in the convolution of the DLA spectrum reads

\begin{equation}
v^{\mathrm{spect}}_{\alpha\beta}[Q]=\sum_{n=1}^Q\frac{U_{Q-n}}{U_Q}\eta_n-\left(\delta E^{\mathrm{spect}}_{\alpha\beta}[Q]\right)^2
\end{equation}

\noindent with

\begin{equation}
\eta_n=\sum_{m=1}^{n-1}\phi_m(D)\phi_{n-m}(D)+n\left(\phi_n\left(D^2\right)+O_n\right),
\end{equation}

\noindent where $\phi_m(D)$ is defined by Eq. (\ref{phim}),

\begin{equation}
\phi_m\left(D^2\right)=-\sum_{s\in\sigma}\left(-X_s\right)^mg_s\left(D_s^{\alpha\beta}\right)^2
\end{equation}

\noindent and

\begin{equation}
O_n=-\sum_{s\in\sigma}\left(-X_s\right)^ng_s~\Delta_s^{(\alpha\beta)}.
\end{equation}

\begin{figure}
\vspace{10mm}
\begin{center}
\includegraphics[width=10cm]{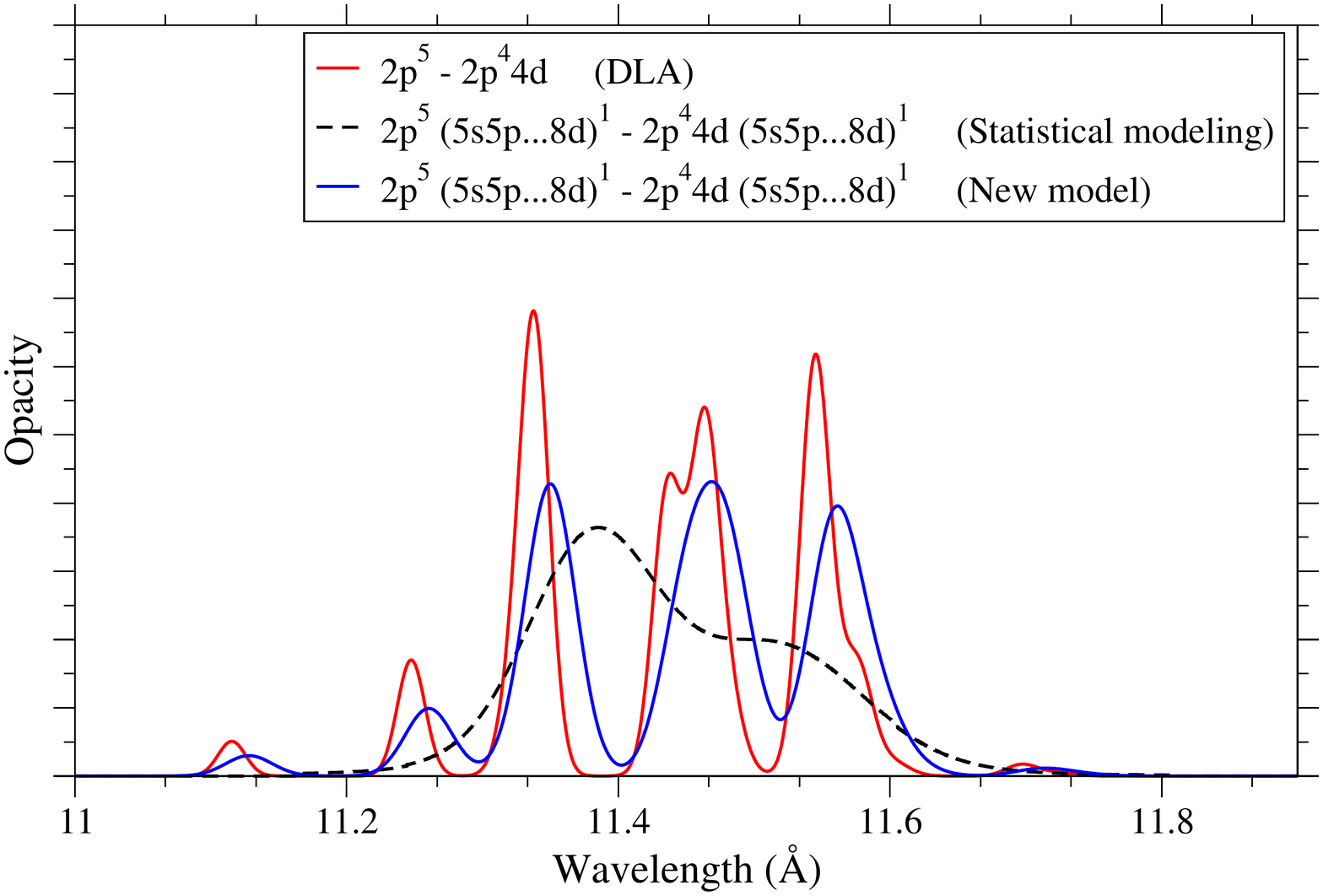}
\vspace{5mm}
\end{center}
\caption{(Color online) Opacity of transition array 2p$^5\rightarrow$ 2p$^4$ 4d$^1$ calculated by SCO-RCG without any spectator (DLA), with a spectator electron $n\ell^1$ (from 5s to 8d) following the procedure described in Sec. \ref{proce} and using a statistical modeling.\label{fig1}}
\end{figure}

\begin{figure}
\vspace{10mm}
\begin{center}
\includegraphics[width=10cm]{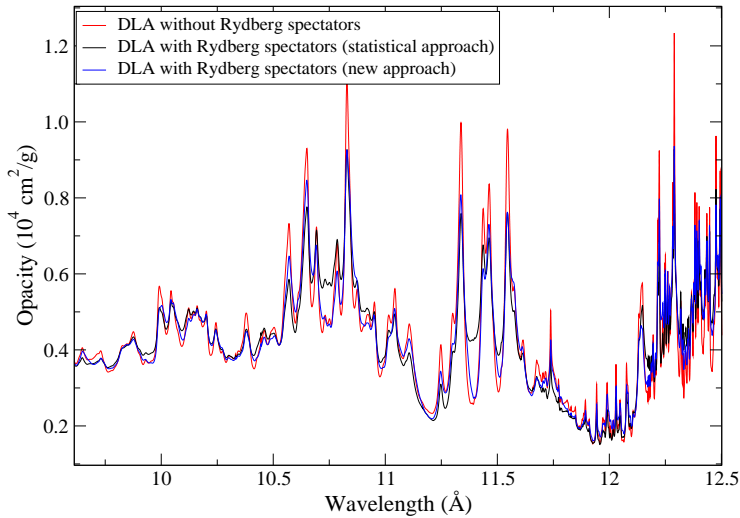}
\vspace{5mm}
\end{center}
\caption{(Color online) Opacity of transition array 2p$^5\rightarrow$ 2p$^4$ 4d$^1$ calculated by SCO-RCG without any spectator and within the previous and new modelings of Rydberg spectators.\label{figure_loisel2014_quad}}
\end{figure}

\noindent If $\alpha\in\sigma$ or $\beta\in\sigma$, then a statistical calculation is performed. We can see that the new modeling fills the gap much less than the previous approach (see Figs. \ref{fig1} and \ref{figure_loisel2014_quad}). As expected, the structure is now close to the resonance one and slightly shifted. We can also observe that the opacity levels of the three substructures are different than the ones of the resonance.

\begin{figure}
\vspace{10mm}
\begin{center}
\includegraphics[width=10cm]{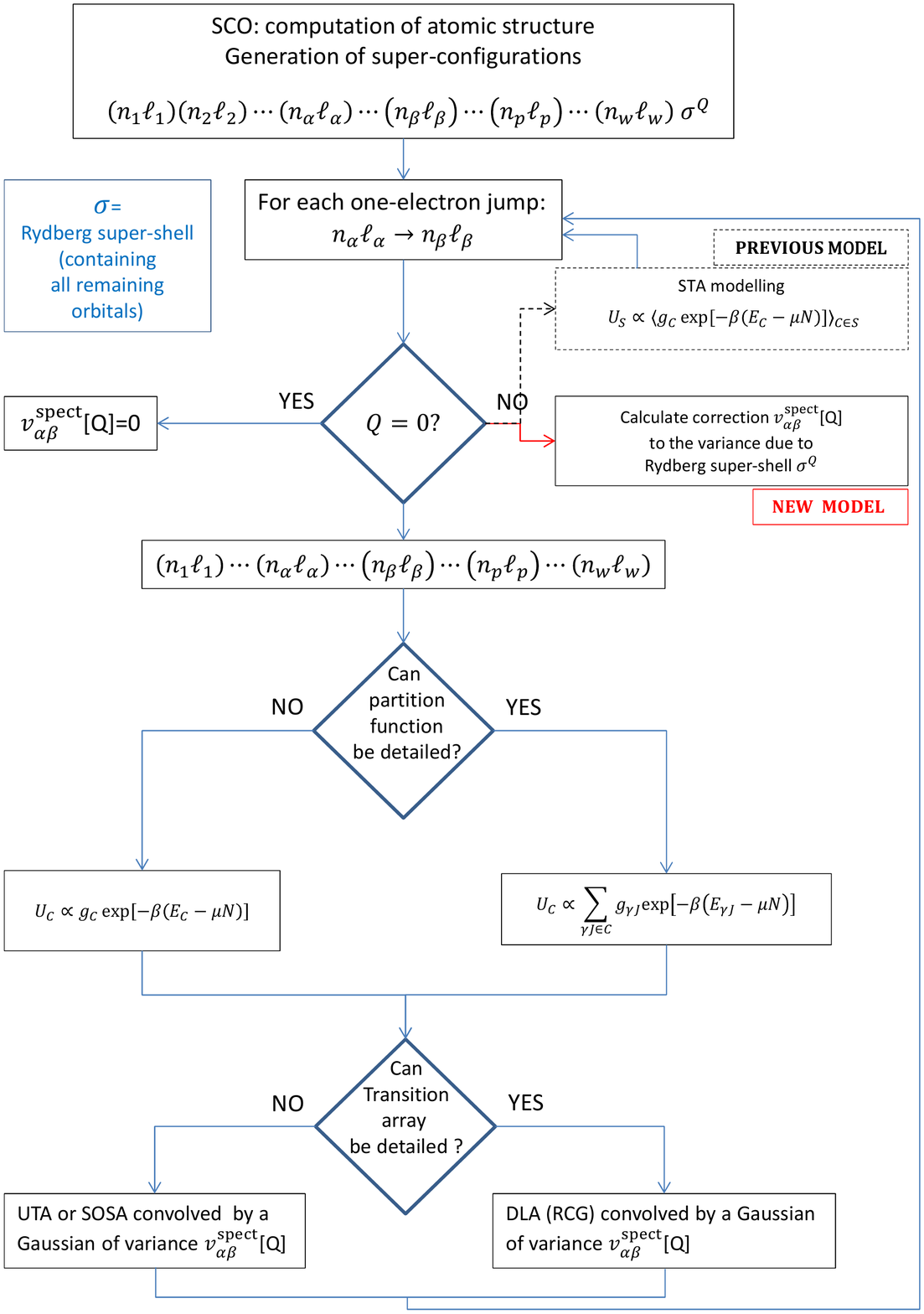}
\vspace{5mm}
\end{center}
\caption{(Color online) Algorithm of the SCO-RCG code.\label{algo}}
\end{figure}

The algorithm of the SCO-RCG code including the modeling of Rydberg super-shell $\sigma^Q$ is described in Fig. \ref{algo}.

\subsection{SWAP approximation}

In the SWAP (Statistical Weight Approximation) limit, \emph{i.e.} when temperature tends to infinity, one has $X_s\equiv 1$ and $U_Q=\left(\begin{array}{c}G\\Q\end{array}\right)$ where $G=\sum_{s\in\sigma}g_s$ is the total degeneracy of super-shell $\sigma$. After simple algebraic manipulation, one finds

\begin{equation}
\delta E^{\mathrm{spect}}_{\alpha\beta}=\frac{Q}{G}\sum_{s\in\sigma}g_sD_s^{\alpha\beta}
\end{equation}

\noindent for the shift of the detailed transition array $\alpha\rightarrow\beta$ and

\begin{equation}
v^{\mathrm{spect}}_{\alpha\beta}[Q]=\frac{Q(G-Q)}{G^2(G-1)}\left(\sum_{s\in\sigma}g_sD_s^{\alpha\beta}\right)^2+\frac{Q(G-Q)}{G(G-1)}\sum_{s\in\sigma}g_s\left(\Delta_s^{(\alpha\beta)}+\left(D_s^{\alpha\beta}\right)^2\right)
\end{equation}

\noindent for the contribution to the variance. It can be checked in table \ref{tab:b} that even at $T$=182 eV, the SWAP approximation for the correction to the shift and broadening due to Rydberg spectators is rather close to their exact values.

\begin{table*}[t]
\begin{center}
\begin{tabular}{|c|c|c|c|}\hline
Transition & Total variance (eV$^2$) & Contribution $v^{\mathrm{spect}}_{\alpha\beta}[Q]$ of $\sigma^1$ (eV$^2$) & $v^{\mathrm{spect}}_{\alpha\beta}[Q]$ in SWAP (eV$^2$) \\ \hline
2s $\rightarrow$ 3p & 0.3132 & 6.42 10$^{-4}$ & 5.28 10$^{-4}$ \\ \hline
2s $\rightarrow$ 4p & 0.3518 & 2.64 10$^{-3}$ & 2.29 10$^{-3}$ \\\hline
2p $\rightarrow$ 3s & 0.1312 & 6.25 10$^{-4}$ & 5.15 10$^{-4}$ \\\hline
2p $\rightarrow$ 3d & 0.1129 & 6.22 10$^{-4}$ & 5.05 10$^{-4}$ \\\hline
2p $\rightarrow$ 4s & 0.1530 & 2.34 10$^{-3}$ & 2.04 10$^{-3}$ \\\hline
2p $\rightarrow$ 4d & 0.1419 & 2.61 10$^{-3}$ & 2.24 10$^{-3}$ \\\hline
\end{tabular}
\caption{Super-configuration (1s)$^2$(2s)$^2$(2p)$^5\sigma^1$ of iron at $T=$182 eV and $N_e$=3.1~10$^{22}$ cm$^{-3}$.}\label{tab:b}
\end{center}
\end{table*}

\subsection{Reducing the statistical part}

\begin{figure}
\vspace{10mm}
\begin{center}
\includegraphics[width=10cm]{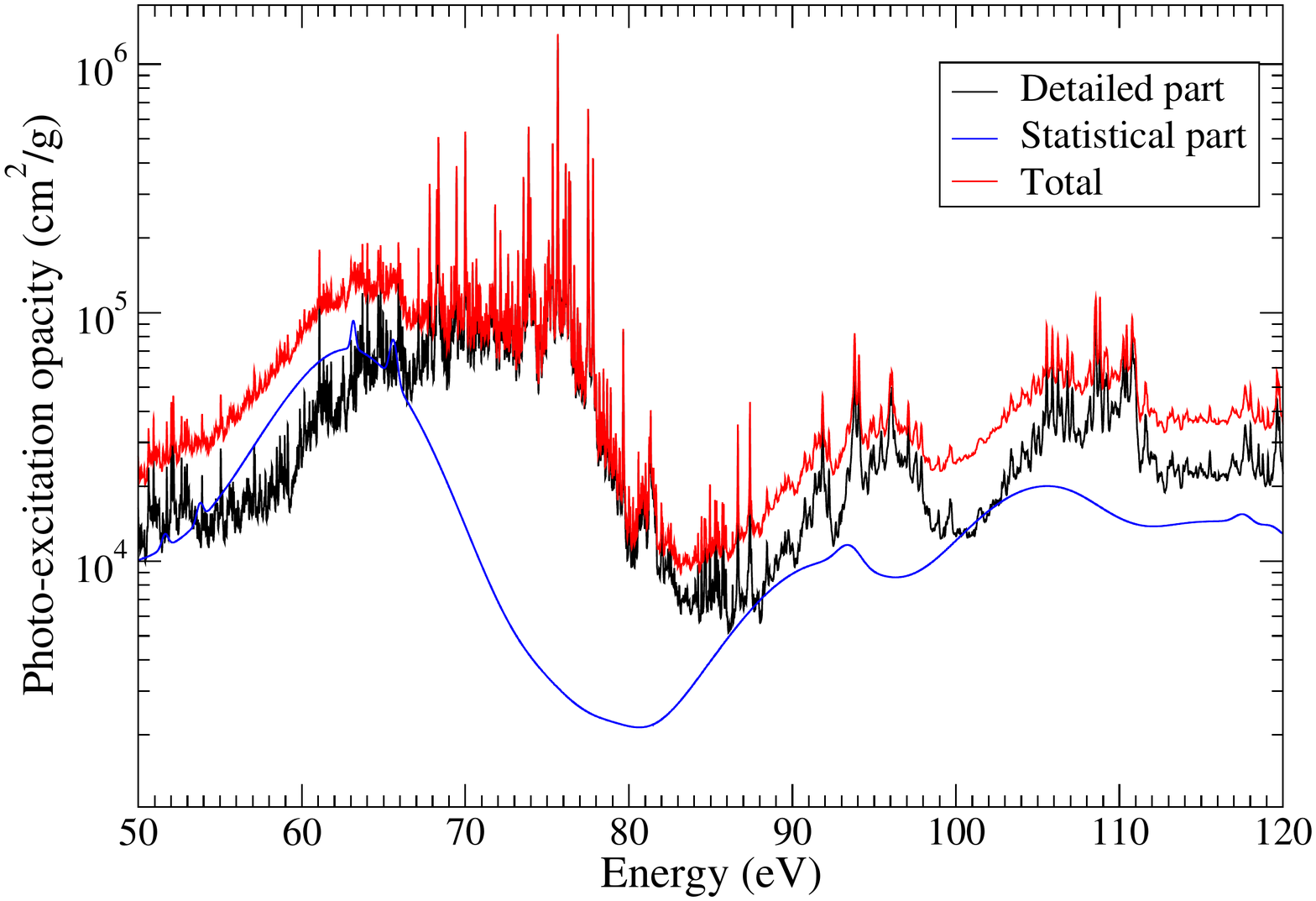}
\vspace{5mm}
\end{center}
\caption{(Color online) Detailed part, statistical part and total photo-excitation opacity calculated by SCO-RCG within the previous modeling of Rydberg spectators in the conditions of Da Silva's experiment ($T$=25 eV and $\rho$=0.008 g/cm$^3$) \cite{DASILVA92}.\label{dasilva_prev}}
\end{figure}

\begin{figure}
\vspace{10mm}
\begin{center}
\includegraphics[width=10cm]{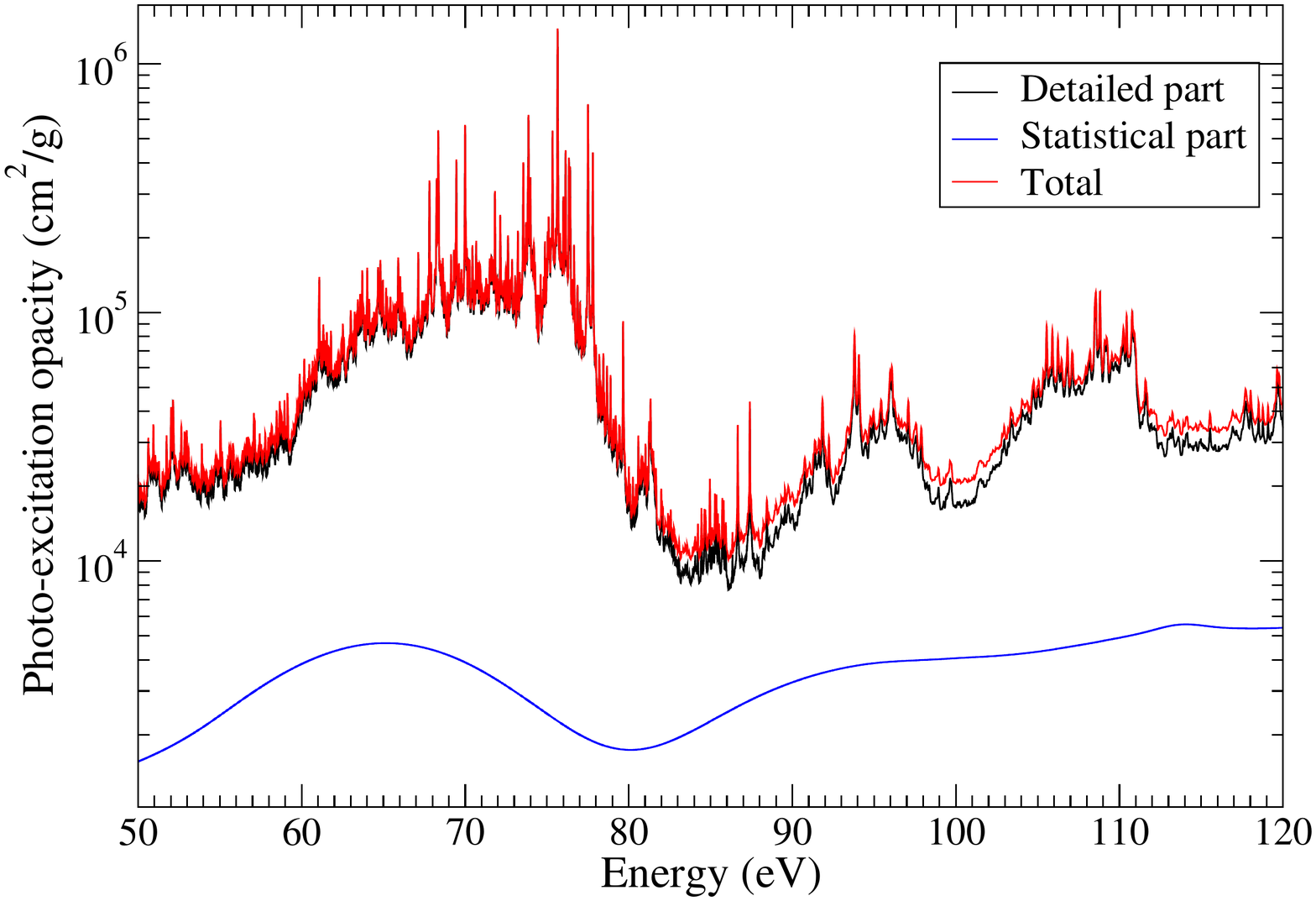}
\vspace{5mm}
\end{center}
\caption{(Color online) Detailed part, statistical part and total photo-excitation opacity calculated by SCO-RCG within the new modeling of Rydberg spectators in the conditions of Da Silva's experiment ($T$=25 eV and $\rho$=0.008 g/cm$^3$).\label{dasilva_ryd}}
\end{figure}

\begin{figure}
\vspace{10mm}
\begin{center}
\includegraphics[width=10cm]{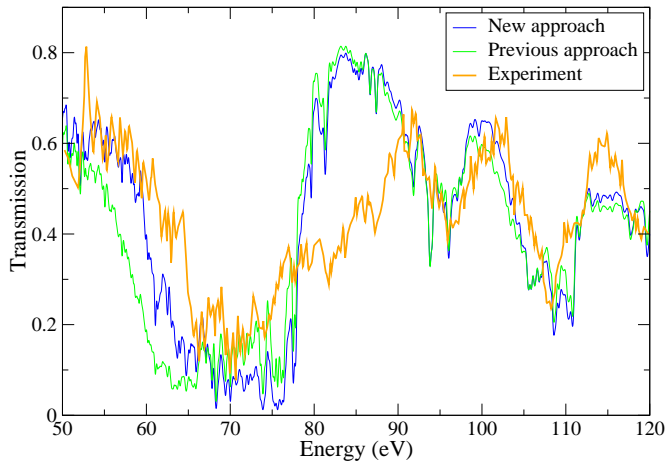}
\vspace{5mm}
\end{center}
\caption{(Color online) Comparison between the experimental spectrum measured by Da Silva \emph{et al.} and transmissions calculated by SCO-RCG within the previous and new modelings of Rydberg spectators in the conditions of Da Silva's experiment ($T$=25 eV and $\rho$=0.008 g/cm$^3$) \cite{DASILVA92}.\label{dasilva}}
\end{figure}

As explained in section \ref{scorcg}, in SCO-RCG, the total photo-excitation opacity consists in two parts: a detailed part, and a statistical part (which is expected to be much smaller than the detailed one). We can see for the case of Da Silva experiment \cite{DASILVA92} at $T$=25 eV and $\rho$=0.008 g.cm$^{-3}$ that the statistical part of the calculation has been significantly reduced within the new modeling of Rydberg spectators (see Figs. \ref{dasilva_prev} and \ref{dasilva_ryd}). The main $\Delta n$=0 (3-3) structure around $h\nu$=70 eV is now closer to the experimental spectrum, although there are still some discrepancies (see Fig. \ref{dasilva}).

\clearpage

\subsection{Impact on the Rosseland mean}

\begin{figure}
\vspace{10mm}
\begin{center}
\includegraphics[width=10cm]{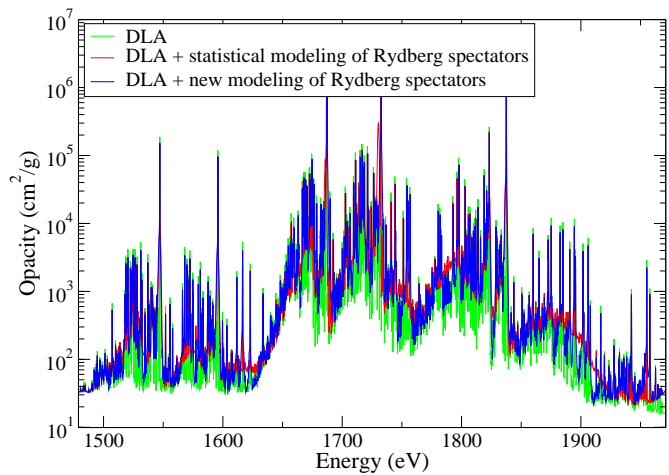}
\vspace{5mm}
\end{center}
\caption{(Color online) Bromine opacity calculated by SCO-RCG: pure DLA (no Rydberg spectators), previous and new modeling of Rydberg spectators.\label{brome}}
\end{figure}

\begin{figure}
\vspace{10mm}
\begin{center}
\includegraphics[width=10cm]{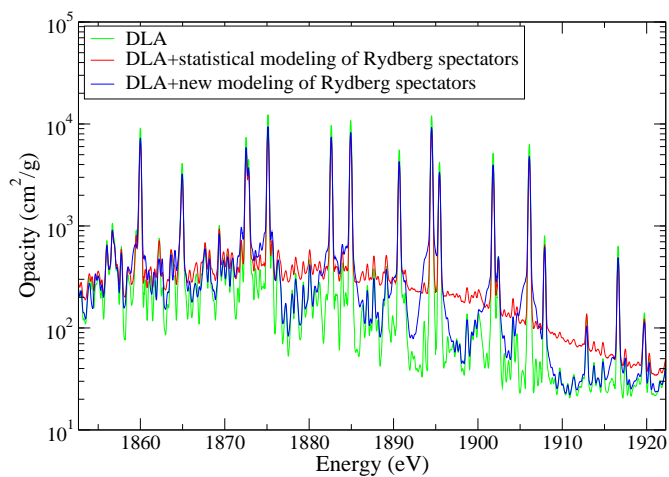}
\vspace{5mm}
\end{center}
\caption{(Color online) Detail of Fig. \ref{brome}.\label{brome_zoom}}
\end{figure}

The new modeling may have a significant impact on the Rosseland mean, which is very sensitive to hollows and windows in the spectrum, and therefore to the line wings. Table \ref{tab:c} indicates the contribution of spectral band [1.5 - 2 keV] to Rosseland opacity, \emph{i.e.}

\begin{equation}
\kappa_R\left[h\nu_1 - h\nu_2\right]=\frac{\int_{h\nu_1}^{h\nu_2}\frac{\partial B_{\nu}(T)}{\partial T}d\nu}{\int_{h\nu_1}^{h\nu_2}\frac{\partial B_{\nu}(T)}{\partial T}\frac{1}{\kappa(h\nu)}d\nu}
\end{equation}

\noindent with $h\nu_1$=1.5 keV and $h\nu_2$=2 keV, for a bromine plasma at $T$=200 eV and $\rho$=0.01 g.cm$^{-3}$ calculated by SCO-RCG within three options: pure DLA (no Rydberg spectators), previous and new modeling of Rydberg spectators (Figs. \ref{brome} and \ref{brome_zoom}). It is clear that omitting the Rydberg electrons is not acceptable and that the differences between the values corresponding to the previous and new approaches are mostly due do the fact that the new model, which can be viewed as a perturbation to the pure DLA computation, fills the gaps between lines in a much less pronounced way than the previous statistical approach.

\begin{table*}[t]
\begin{center}
\begin{tabular}{|c|c|c|c|}\hline
Transition & DLA             & DLA              & DLA \\
           & (no Ryd. spec.) & + stat. modeling & + new modeling \\
           &                 & of Ryd. spec.    & of Ryd. spec. \\ \hline
$\kappa_R\left[1.5 - 2~\mathrm{keV}\right]$ (cm$^2$/g) & 901.27 & 1227.52 & 1129.42 \\ \hline
\end{tabular}
\caption{Contribution of the frequency band [1.5 - 2 keV] to the Rosseland mean with the previous statistical model and the new approach compared to the calculation without DLA spectators. The Rydberg super-shell contains subshells 5s to 15t.}\label{tab:c}
\end{center}
\end{table*}

\subsection{Remark concerning autoionizing states} \label{rem}

\begin{figure}
\vspace{10mm}
\begin{center}
\includegraphics[width=10cm,angle=-90]{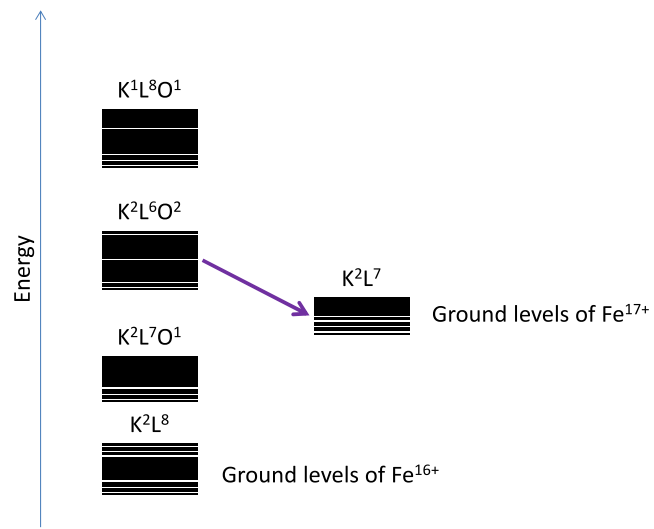}
\vspace{10mm}
\end{center}
\caption{(Color online) Autoionizing complexes in the case of iron.\label{figure_ppt2}}
\end{figure}

Some autoionizing states \cite{BAUCHE09,POIRIER88} may lead to an additional broadening of the absorption structures (not taken into account in SCO-RCG). For instance, the super-configuration K$^2$L$^6$O$^2$ of ion Fe$^{16+}$ lies above the fundamental super-configuration K$^2$L$^7$ of ion Fe$^{17+}$ (see Fig. \ref{figure_ppt2}). Therefore, considering only one electron in the Rydberg super-shell (which, in the present case, starts at orbital 5s), this effect is not important. However, for two or more electrons in the Rydberg super-shell, this broadening mechanism must be taken into account. Work is in progress to include it in SCO-RCG. 

\subsection{Prescription for the inclusion of Rydberg spectators in opacity codes}

Let us consider the super-configuration

\begin{equation}
\left\{\begin{array}{l}
S_Q=C~\sigma^Q\\
C=\left(n_1\ell_1\right)^{N_1}\left(n_2\ell_2\right)^{N_2}\left(n_3\ell_3\right)^{N_3}\cdots\left(n_w\ell_w\right)^{N_w},
\end{array}\right.
\end{equation}

\noindent $\sigma$ representing the Rydberg super-shell. One could consider convolving the detailed transition array $C\rightarrow C'$ by a linear combination of profiles

\begin{equation}
\psi_{S_Q\rightarrow S'_Q}(h\nu)=V\left(h\nu-\delta E_Q,a^{C\rightarrow C'},v^{C\rightarrow C'}+v_Q\right)
\end{equation}

\noindent where the shift reads

\begin{equation}
\delta E_Q=\sum_{n=1}^Q\frac{U_{Q-n}(g)}{U_Q(g)}\phi_n(D),
\end{equation}

\noindent the correction to the variance

\begin{equation}
v_Q=\sum_{n=1}^Q\frac{U_{Q-n}(g)}{U_Q(g)}\eta_n-\left(\delta E_Q\right)^2
\end{equation}

\noindent and

\begin{equation}
\mathcal{V}(x,a,\nu)=\frac{a}{\pi\sqrt{2\pi v}}\int_{-\infty}^{\infty}\frac{e^{-\frac{y^2}{2v}}}{a^2+(x-y)^2}dy
\end{equation}

\noindent is the Voigt function.

\section{Conclusion}\label{conc}

Spectator electrons in high-lying states may be important in X-ray absorption or emission spectra in LTE or NLTE hot plasmas. We proposed a new model, based on the additivity of the variances from active and spectator electrons in the STA theory. This approach, inspired from the Partially-Resolved-Transition-Array model, consists in a reduced detailed-line-accounting calculation, omitting these Rydberg electrons, the effect of the latter being included through an additional shift and broadening of the lines, expressed in terms of canonical partition functions. The resulting method can be used in any DLA opacity code. 

\section{Appendix: The formula of Goldberg and Rozsnyai}

In their work \cite{GOLDBERG86}, Goldberg and Rozsnyai neglected the effect of terms as well as explicit electron-electron interactions in the Boltzmann factors. Their expression was obtained from the formula

\begin{equation}
v_{\mathrm{dielectronic}}=\sum_{\scriptsize\begin{array}{c}p~\mathrm{Rydberg}\\p\ne\alpha,\beta\end{array}}\left(V_{p\beta}-V_{p\alpha}\right)^2\left[\langle N_p^2\rangle-\langle N_p\rangle^2\right],
\end{equation}

\noindent where the average value of a function $\langle A\left(\vec{N}\right)\rangle$, with $\vec{N}=\left(N_1,N_2,N_3,\cdots, N_w\right)$, is obtained as 

\begin{equation}\label{ave}
\langle A\left(\vec{N}\right)\rangle=\frac{\displaystyle\sum_{N_1=0}^{g_1}\sum_{N_2=0}^{g_2}\cdots\sum_{N_w=0}^{g_w}\left\{\prod_{i=1}^w\bin{g_i}{N_i}X_i^{N_i}\right\}N_{\alpha}\left(g_{\beta}-N_{\beta}\right)A\left(\vec{N}\right)}{\displaystyle\sum_{N_1=0}^{g_1}\sum_{N_2=0}^{g_2}\cdots\sum_{N_w=0}^{g_w}\left\{\prod_{i=1}^w\bin{g_i}{N_i}X_i^{N_i}\right\}N_{\alpha}\left(g_{\beta}-N_{\beta}\right)}
\end{equation}

\noindent with $X_i=e^{-\beta\left(\epsilon_i-\mu\right)}$. This is a grand-canonical evaluation of the averages, \emph{i.e.} without any assumption concerning the number of electrons. Applying formula (\ref{ave}) to $A\left(\vec{N}\right)=N_p$ and $A\left(\vec{N}\right)=N_p^2$ and using elementary binomial relations, one finds

\begin{equation}
v_{\mathrm{dielectronic}}=\sum_{\scriptsize\begin{array}{c}p~\mathrm{Rydberg}\\p\ne\alpha,\beta\end{array}}\left(V_{p\beta}-V_{p\alpha}\right)^2\bar{N}_p\left(1-\frac{\bar{N}_p}{g_p}\right),
\end{equation}

\noindent where $\bar{N}_p=g_pX_p/\left(1+X_p\right)$ is the average population of orbital $p$.

\section{Acknowldgements}

The authors would like to thank J. Bailey, T. Nagayama and G. Loisel for stimulating the present study.

\end{document}